\documentclass[a4paper,fleqn,usenatbib]{mnras}

\usepackage[T1]{fontenc}
\usepackage{ae,aecompl}

\usepackage{graphicx}	
\usepackage{amsmath}	
\usepackage{amssymb}	

\usepackage{subcaption} 
\usepackage{placeins}
\usepackage{float}

\usepackage{mathrsfs}
\usepackage[flushleft]{threeparttable}
\usepackage{verbatim}
\usepackage{multicol}

\usepackage{newtxtext,newtxmath}
\usepackage[normalem]{ulem}

\defcitealias{Gaulme2016}{\scshape G16}
\newcommand*{\dittoclosing}{--\,\raisebox{-0.5ex}{''}\,--}

\title[The accuracy of asteroseismology]{Establishing the accuracy of asteroseismic mass and radius estimates of giant stars \\II. Revised stellar masses and radii for KIC\,8430105}

\author[J. S. Thomsen et al.]{J. S. Thomsen,$^{1}$\thanks{E-mail: jet@phys.au.dk}
K. Brogaard,$^{1}$
T. Arentoft,$^{1}$
D. Slumstrup,$^{2,1}$
M. N. Lund,$^{1}$
\newauthor 
F. Grundahl,$^{1}$
A. Miglio,$^{3, 4, 1}$
J. Jessen-Hansen,$^{1}$
S. Frandsen$^{1}$
\\
$^{1}$Stellar Astrophysics Centre, Department of Physics and Astronomy, Aarhus University, Ny Munkegade 120, 8000 Aarhus C, Denmark\\
$^{2}$European Southern Observatory, Alonso de Cordova 3107, Vitacura, Santiago, Chile\\
$^{3}$Dipartimento di Fisica e Astronomia, Università di Bologna, Via Zamboni, 33 - 40126 Bologna, Italia\\
$^{4}$INAF-Osservatorio di Astrofisica e Scienza dello Spazio di Bologna, via Gobetti 93/3, I-40129 Bologna, Italy
}

\date{Accepted XXX. Received YYY; in original form ZZZ}

\pubyear{2017}

\begin{document}
\label{firstpage}
\pagerange{\pageref{firstpage}--\pageref{lastpage}}
\maketitle

\begin{abstract}
Asteroseismic scaling relations can provide high-precision measurements of mass and radius for red giant (RG) stars displaying solar-like oscillations. Their accuracy can be validated and potentially improved using independent and accurate observations of mass, radius, effective temperature and metallicity. We seek to achieve this using long period SB2 eclipsing binaries hosting oscillating RGs.
We explore KIC\,8430105, for which a previous study found significant asteroseismic overestimation of mass and radius when compared with eclipsing binary measurements. We measured dynamical masses and radii for both components to be significantly lower than previously established, increasing the discrepancy between asteroseismic and dynamical measurements.
Our dynamical measurements of the RG component were compared to corresponding measurements of mass and radius using asteroseismic scaling relations. Uncorrected scaling relations overestimated the mass of the RG by $26\%$, the radius by $11\%$, and the average density by $7\%$, in agreement with studies for other systems. However, using a theoretical correction to $\Delta\nu$, we managed to obtain an asteroseismic average density that is $1\sigma$ consistent with our dynamical result.
We obtained several measurements of $\nu_{\rm max}$ that are not fully consistent. With $\nu_{\rm max} = 76.78 \pm 0.81 \mu$Hz, the $\Delta\nu$ correction provided $2\sigma$ consistent mass and radius for the giant.
The age of the system was estimated to be $3.7\pm0.4$ Gyr.
\end{abstract}




\section{Introduction}



The asteroseismic scaling relations can be a powerful tool for establishing, fundamental stellar parameters, specifically the mass and the radius. However, as mentioned by \citet{Brogaard2018} (Hereafter Paper I), they must be empirically verified before proper use in ensemble observations of evolved stars, and calibrated if necessary.

Paper I described the overall basis of using detached eclipsing binaries (dEBs) that are also spectroscopic double-lined binaries (SB2) to establish the accuracy of asteroseismic mass and radius. These systems are unique as it is possible to obtain model-independent and high-precision radius, mass, effective temperature and metallicity of both stellar components. When one of the stellar components also shows solar-like oscillations, the stellar parameters determined with dEB analysis can be directly compared with asteroseismic scaling relation measurements. 

The scaling relations in equation form are \citep{Brown1991,Kjeldsen1995,Belkacem2011}:

\begin{eqnarray}\label{eq:01}
\frac{\Delta \nu}{\Delta \nu _{\odot}} & \simeq & f_{\Delta \nu}\left(\frac{\rho}{\rho_{\odot}}\right)^{1/2},\\
\label{eq:02}
\frac{\nu _{\mathrm{max}}}{\nu _{\mathrm{max,}\odot}} & \simeq & f_{\nu _{\mathrm{max}}} \frac{g}{g_{\odot}}\left(\frac{T_{\mathrm{eff}}}{T_{\mathrm{eff,}\odot}}\right)^{-1/2}.
\end{eqnarray}
Here, $\rho$, $g$, and $T_{\rm eff}$ are the mean density, surface gravity, and effective temperature. $\Delta\nu$ is the asteroseismic large frequency spacing, and $\nu_{\rm max}$ is the asteroseismic frequency of maximum power. We have adopted the notation of \citet{Sharma2016} that includes the correcting terms $f_{\Delta \nu}$ and $f_{\nu _{\rm max}}$. We adopt the solar reference values of $\Delta \nu _{\odot} = 134.9 \mu$Hz and $\nu _{\rm max,\odot} = 3090 \mu$Hz following \citet{Handberg2017}. Rearranging these expressions gives scaling relations for mass and radius: 

\begin{eqnarray}\label{eq:03}
\frac{M}{\mathrm{M}_\odot} & \simeq & \left(\frac{\nu _{\mathrm{max}}}{f_{\nu _{\mathrm{max}}}\nu _{\mathrm{max,}\odot}}\right)^3 \left(\frac{\Delta \nu}{f_{\Delta \nu}\Delta \nu _{\odot}}\right)^{-4} \left(\frac{T_{\mathrm{eff}}}{T_{\mathrm{eff,}\odot}}\right)^{3/2},\\
\label{eq:04}
\frac{R}{\mathrm{R}_\odot} & \simeq & \left(\frac{\nu _{\mathrm{max}}}{f_{\nu _{\mathrm{max}}}\nu _{\mathrm{max,}\odot}}\right) \left(\frac{\Delta \nu}{f_{\Delta \nu}\Delta \nu _{\odot}}\right)^{-2} \left(\frac{T_{\mathrm{eff}}}{T_{\mathrm{eff,}\odot}}\right)^{1/2}. 
\end{eqnarray}
Paper I investigated three SB2 eclipsing binaries with an oscillating giant component. They found systematic overestimation of mass and radius when applying no correction to $\Delta \nu$. When applying a theoretical correction $f_{\Delta\nu}$ from \citet{Rodrigues2017}, they achieved general agreement between asteroseismic and dynamically derived masses.

\citet{Gaulme2016} (Hereafter \citetalias{Gaulme2016}) analysed a sample of 10 SB2 eclipsing binaries with detectable oscillations in one component, including the three systems also investigated in Paper I. They explored several different $\Delta\nu$ corrections and concluded that corrected masses and radii were overestimated in general. However, as mentioned by Paper I this does not hold for all the individual systems in the sample, with some showing agreement with corrected scaling relations and others deviating significantly. Multiple observational explanations for this were not ruled out. This includes zero-point issues of the measurements of the radial velocities (RVs) and uncertainty underestimation of those RVs and the dynamically derived parameters. It also might include temperature overestimation due to neglecting the contribution of the secondary component to the spectroscopic continuum when co-adding stellar spectra.

The goal of this paper is to independently investigate one of the systems, KIC\,8430105, from the \citetalias{Gaulme2016} sample that had a large asteroseismic overestimation of mass and radius even when using corrections. This system has a significant luminosity difference between the two components, with the secondary main-sequence (MS) component contributing less than $2\%$ of the total luminosity. This suggests that temperature overestimation from neglecting the continuum of the secondary is unlikely to be the primary cause of asteroseismic overestimation. However it also puts significant requirements for high signal-to-noise ratio (S/N), high resolution, well-calibrated spectra in order to be able to measure precise RVs for both components, which we provide with this paper.

The system in question, KIC\,8430105, has been modelled with asteroseismology in \citet{buldgen2019} using the individual radial-mode frequencies. They used the \emph{AIMS} software \citep{rendle2019} and a two-term surface correction from \citet{ballgizon2014}. This method does not depend on the frequency of maximum power $\nu_{\rm max}$. \citet{Joergensen2020} also investigated KIC\,8430105 with \emph{AIMS} and multiple different surface corrections. \citet{Li2022} compared literature values of mass and radius from eclipsing binary studies of 5 other systems with their individual frequency modelling using two-term \citet{ballgizon2014} surface corrections.

For the scaling relations, we will focus on the $\Delta\nu$ corrections by \citet{Rodrigues2017}. Paper I suggests that this correction might be sufficient. We will also compare with a proposed theoretical $\nu_{\rm max}$ correction by \citet{Viani2017} which couples the corresponding scaling relation, Eq.~\eqref{eq:02}, to the stellar mean molecular weight and adiabatic exponent.

In Sects.~\ref{sec:obs} -- \ref{sec:binary} we present observations and precise measurements of masses, radii, effective temperatures and metallicities of KIC\,8430105. In Sect.~\ref{sec:distance} we compare the dynamical radius of the RG with radius estimated using the Gaia parallax and photometry along with the spectroscopic temperature determined in this study. In Sect.~\ref{sec:age} we provide an age estimate for the system using the dynamically determined parameters. In Sect.~\ref{sec:compare} we compare the dynamical parameters to asteroseismic predictions using scaling relations, corrections and modelling. Further, we discuss whether the findings of a large asteroseismic overestimation of mass and radius by \citetalias{Gaulme2016} is correct for this system, and if so, what this suggests. %
Finally, in Sect.~\ref{sec:conclusion}, we summarise, conclude and outline paths to improve upon the accuracy level of asteroseismology of giants in future studies.
\section{Observations and observables}
\label{sec:obs}
%
%
%
%
%
%
%
%
\subsection{Photometry}
Three processed light curves based on data from the \emph{Kepler} mission \citep{Borucki2010} was used during separate parts of the analysis. For asteroseismic analysis, light curves were created from \textit{Kepler} target pixel files \citep{jenkins10}, downloaded from the KASOC database\footnote{\url{kasoc.phys.au.dk}}, and generated using the procedure developed by S. Bloemen (private communication) to automatically define pixel masks for high S/N aperture photometry, similar to what was done in Paper I. The extracted light curves were then corrected using the KASOC filter \citep{Handberg2014}. In total, three different filters are attained in this that can each be used to isolate separate parts of the light curve. The process is detailed in \citet{Handberg2014}.

We found that one of the three, the long time-scale filter, was over-fitting the eclipses of KIC\,8403015. This was corrected for by the subsequent eclipse filter before asteroseismic analysis. However, with this knowledge we decided not to use the KASOC filters for the eclipsing binary analysis. As detailed in Appendix.~\ref{sec:tlight}, contamination for this light curve is likely higher than the estimates produced for the \emph{Kepler} simple aperture photometry (SAP) pipeline due to the KASOC pipeline valuing larger apertures on average \citep{Handberg2014}. Therefore, the final eclipsing binary measurements reported were instead produced using the \emph{Kepler} Presearch Data Conditioning light curve \citep[PDCSAP, ][]{PDCSAP_1_2012, PDCSAP_2_2012, PDCSAP_3_2017} for the system, downloaded from MAST\footnote{\url{https://mast.stsci.edu/portal/Mashup/Clients/Mast/Portal.html}}, while the light curve produced using KASOC was only utilized to perform asteroseismic analysis, and to examine the effects of third light on the observed dynamical parameters.

To correct for long-term variations caused by instrumental and astrophysical variability, we performed local polynomial fitting around each of the eclipse epochs followed by removal of most of the out-of-eclipse observations. We estimated the photometric uncertainty by calculating the RMS of the phase-folded light curve within the total eclipse of the MS star.

KIC\,8430105 was also part of the sector 15 2-min cadence TESS \citep{Ricker2014} data, which managed to capture a single instance of both eclipses of the system. We performed a separate eclipsing binary analysis using this light curve. The Lightkurve software \citep{lightkurve2018} was used to perform aperture photometry with the TESS pipeline aperture, and to perform general corrections using principal components analysis. Polynomial fitting around the eclipses was done to remove out-of-eclipse trends, and the light curve was truncated. Photometric uncertainty was ascertained in a similar fashion to the \emph{Kepler} light curve. The photometric precision of the TESS light curve is lower, and it has a high degree of contamination ($\sim 8\%$). Focus in this paper will therefore be on the results produced using the \emph{Kepler} light curve. A detailed investigation of contamination is presented in Appendix~\ref{sec:tlight}.

\subsection{Spectroscopy}
We used spectroscopic follow-up observations obtained with the FIES spectrograph \citep{Telting2014} at the Nordic Optical Telescope (NOT) located at the Observatorio del Roque de los Muchachos on La Palma. The spectral resolution was $R\sim67000$ and the integration time was 1800 seconds. \emph{FIEStool} \citep{FIEStool} was used for data reduction, extraction, wavelength calibration and order merging. To calculate barycentric corrections and barycentric julian dates, we used the software \emph{barycorrpy} described in \citet{kanodia2018}. Table~\ref{tab:rv8430105} in the appendix gives the barycentric julian dates, S/N, radial velocities, and weight applied to the spectrum when used in the spectral separation routine described during Sect.~\ref{sssec:rv}.

\subsubsection{Radial velocity measurements}\label{sssec:rv}
For radial velocity (RV) measurements and separation of the component spectra, we developed a python code \href{https://jsinkbaek.github.io/sb2sep/}{\emph{sb2sep}} based on the IDL spectral separation code used in Paper I. It follows the descriptions of \citet{Gonzalez2006}. It uses the broadening function formulation \citep{Rucinski1999,Rucinski2002} with synthetic spectra from \citet{Coelho2005} for radial velocity determination.

We utilized a large wavelength interval 4500--5825Å for the RV measurements, and then split the whole spectrum into 7 smaller intervals of 265Å independently in order to produce error estimates. These intervals were: 4500--4765, 4765--5030, 5030--5295, 5295--5560, 5560--5825, 5985--6250 and 6575--6840Å. The last two intervals were chosen such that the interstellar Na lines and telluric lines were avoided. We found that using the whole 4500--5825Å interval for the measurement, instead of taking the unweighted average of the smaller intervals, produced a significantly smaller spread of the RVs for the giant component during fitting than the uncertainty estimates from the smaller intervals indicated. A possible explanation of this is either that wavelength dependent systematic small-scale shifts produced deviations between the small interval observations, or that a wavelength dependent S/N was better weighted in the broadening function calculation when using most of the spectrum. We assumed that this affected all spectra equally, and corrected for it by subtracting a fixed value of $\rm 45$ m/s from all RV uncertainty estimates of both components. We used the reduced $\chi^2$ of the RVs of the giant component in the dynamical solution to select this value.
With this, the mean RV uncertainty was $42$\,m/s and $0.6$\,km/s for the giant and MS component, respectively.

The MS component contributes less than $2\%$ of the light in the \emph{Kepler} photometry. Following Paper I, we manually inspected the broadening functions for the MS component and discarded RV measurements where it could not be identified with certainty. We did not utilize all the available spectra to calculate the separated component spectra, but selected a subset based on both S/N and phase such that the spectra were as equally spaced in RV as possible. Spectra observed within eclipses were also excluded from this calculation.
\subsubsection{Spectral analysis}\label{sec:specanal}
\begin{table}
\centering
\caption{\label{table:stelpar}Atmospheric parameters for the RG in KIC\,8430105.}\begin{tabular}{lc}
\hline
\hline
Quantity    & Value \\
\hline
$T_{\rm eff} (\rm K)$   &   $4990 (30)$ \\
$\rm [FeI/H] (dex)$  &   $-0.46 (1)$ \\
$\rm [FeII/H] (dex)$ &   $-0.50 (4)$ \\
$\rm [\alpha/Fe] (dex)$ &   $0.08 (4)$ \\
$\xi (\rm km s^{-1})$ &     $0.91 (4)$ \\
\hline
\end{tabular}
\end{table}
The separated component spectra were renormalized using the light ratio from the eclipsing binary analysis in Sect.~\ref{sec:binary}. The stellar atmospheric parameters of the RG component in Table~\ref{table:stelpar} were then determined with a classical equivalent-width spectral analysis on the separated component spectra following the method outlined in \citet{Slumstrup2019}. The $\log g$ for the RG component was kept fixed at the dynamically measured value during this analysis. The auxiliary program \emph{Abundance} with SPECTRUM \citep{Gray1994} was used to determine the atmospheric parameters. We used MARCS stellar atmosphere models \citep{Gustafsson2008} assuming LTE, with solar abundances from \citet{Grevesse1998}. The oscillator strengths for each absorption line have been calibrated on a solar spectrum by adjusting the $\log \text{gf}$ until the well-established solar abundances were achieved \citep[for further explanation see][]{Slumstrup2019}. We measured abundances of [Fe/H] and [$\alpha$/Fe], with the latter defined as $\frac{1}{4} \cdot \left( \text{ [Ca/Fe] + [Si/Fe] + [Mg/Fe] + [Ti/Fe] } \right)$. The total flux from the MS component is very low and thus its separated spectrum is noise dominated making a spectral analysis meaningless.

Results in Table~\ref{table:stelpar} has uncertainties that reflect only line-to-line scatter and not systematic uncertainty. For use in later analysis we follow the investigations of \citet{Bruntt2010} and adopt total $1\sigma$ uncertainties of 80 K for the effective temperature and 0.1dex for [Fe/H].
\section{Eclipsing binary analysis}
\label{sec:binary}
For the eclipsing binary analysis, we used the JKTEBOP code \citep{Southworth2004}, which is based on the EBOP program developed by P. Etzel \citep{Etzel1981, Popper1981}. We made use of non-linear limb darkening \citep{Southworth2007}, independent third light measurement \citep{Southworth2010}, simultaneous fitting of the light curve and the measured radial velocities \citep{Southworth2013}, and numerical integration \citep{Southworth2011}. The latter was necessary due to the integration time of {\it Kepler} long cadence photometry (29.4 minutes).

We fitted for these parameters: Orbital period $P$, a reference eclipse time of the giant component $t_G$, central surface brightness ratio $J$, sum of the relative radii $r_{\rm MS}+r_{\rm G}$, ratio of the radii $k=\frac{r_{\rm MS}}{r_{\rm G}}$, orbit inclination $i$, $e \cos \omega$, $e \sin \omega$, a light scale factor, the linear component of quadratic limb darkening for the giant $ld_{a, \rm RG}$, RV semi-amplitudes of the components $K_{\rm G}$ and $K_{\rm MS}$, and a system velocity for both of the components $\gamma_{\rm G}$ and $\gamma_{\rm MS}$. We allow for two system velocities since the stellar components and their spectra are affected differently by gravitational redshift \citep{Einstein1952} and convective blueshift \citep{Gray2009}. 

We fit for a light scale factor despite using a normalized light curve, in order to account for uncertainty on the polynomial normalization both during the best-fit and during subsequent uncertainty estimation. To investigate whether this skewed the best fit, we compared residual plots produced with and without fitting a light scale factor, and found no visual difference in residuals outside of the eclipses. The fitted light scale factor was $(-1.2 \pm 3.3) \cdot 10^{-5}$mag. Not fitting for it produced slightly lower radii of the components, about $10\%$ the size of our estimated $1\sigma$ uncertainty. We note that since the radii would be only slightly lower by not fitting for the light scale factor, this would not change any conclusions of the paper in Sect.~\ref{sec:conclusion}.

The TESS light curve covered less than a full orbit of the system, meaning that the orbital period cannot be directly determined from the photometry. For the model based on the TESS light curve, we therefore chose not to fit for the orbital period, and instead used the value found with the \emph{Kepler} light curve as a fixed parameter.

We used a quadratic limb darkening law with coefficients linearly interpolated from tabulations by \citet{Claret2011} for the $K_p$ bandpass and from \citet{claret2017} for the TESS bandpass. We ran JKTEBOP iteratively, starting with limb darkening coefficients from first guesses and then using our spectroscopic $T_{\rm eff}$ for the RG and $\log g$ fixed from the binary solution. By reproducing the dynamically determined light ratio in the \emph{Kepler} passband using Planck functions with the $T_{\rm eff}$ of the giant and the ratio of the radii, an estimate of $T_{\rm eff}$ for the MS component was obtained. New limb darkening coefficients were then interpolated using these $T_{\rm eff}$ and log$g$ values to be used in the next JKTEBOP solution. To improve self-consistency and reduce model dependency, we decided to fit for one of the two limb darkening coefficients for the giant. Here we chose the linear coefficient. The obtained radius of the giant was smaller by $0.3\sigma$ when using purely interpolated coefficients and larger by $0.5\sigma$ when instead fitting for the non-linear coefficient. When fitting for both limb darkening coefficients for the giant, the non-linear term is found to be unlikely at $0.016$, ${\sim}10$ times smaller than the tabulated value, indicating that we are not able to fit for both limb darkening coefficients of the giant simultaneously. These tests indicate the level of model-independent accuracy available. From the perspective of only our modelling of the light curve, it would be reasonable to expect a bias of up to $\sim 0.5\sigma$ on the radius due to dependence on stellar models through limb darkening coefficients.

Gravity darkening coefficients were fixed to 0.0 for both components. We compared with results obtained using coefficients taken from \citet{Claret2011} and found no discernible difference in the fitted parameters. Reflection and deformation approximations were disabled since we normalized the light curve outside of eclipses (reflection coefficients fixed at 0.0, mass ratio fixed at -1.0 in JKTEBOP). When instead using the mass ratio from the radial velocity fit, we found no significant change in measured radii. In Appendix~\ref{sec:forward} we compare the JKTEBOP bi-axial ellipsoid approximation with the more advanced models of the PHOEBE 2 code \citep{conroy2020}. The effect on dynamical parameters from normalizing the light curve using polynomial fitting around the eclipses is also investigated in Appendix~\ref{sec:forward}. We found it unlikely that this method introduced a significant bias for the measured stellar radii of KIC\,8430105.

We performed 4.0$\sigma$ clipping of the photometry. This primarily excluded some data-points within the annular eclipse of the giant that we suspected to associate with the MS star transiting stellar spots on the giant.

We obtained reduced $\chi ^2$ values of 0.92 for the total fit, 0.92 for the light curve fit, 0.95 for the RV model of the giant, and 1.66 for the RV model of the MS star. To compare, we also utilized the build-in JKTEBOP option '\emph{chif}' to iteratively adjust uncertainties in order to match reduced $\chi ^2 \sim 1$. This did not produce changes anywhere near significance for any parameters. As an example, the change found for the radius of the giant was on the order of $ 2 \%$ the size of the estimated $1\sigma$ uncertainty.

\begin{figure}
   \centering
    \includegraphics[width=\columnwidth]{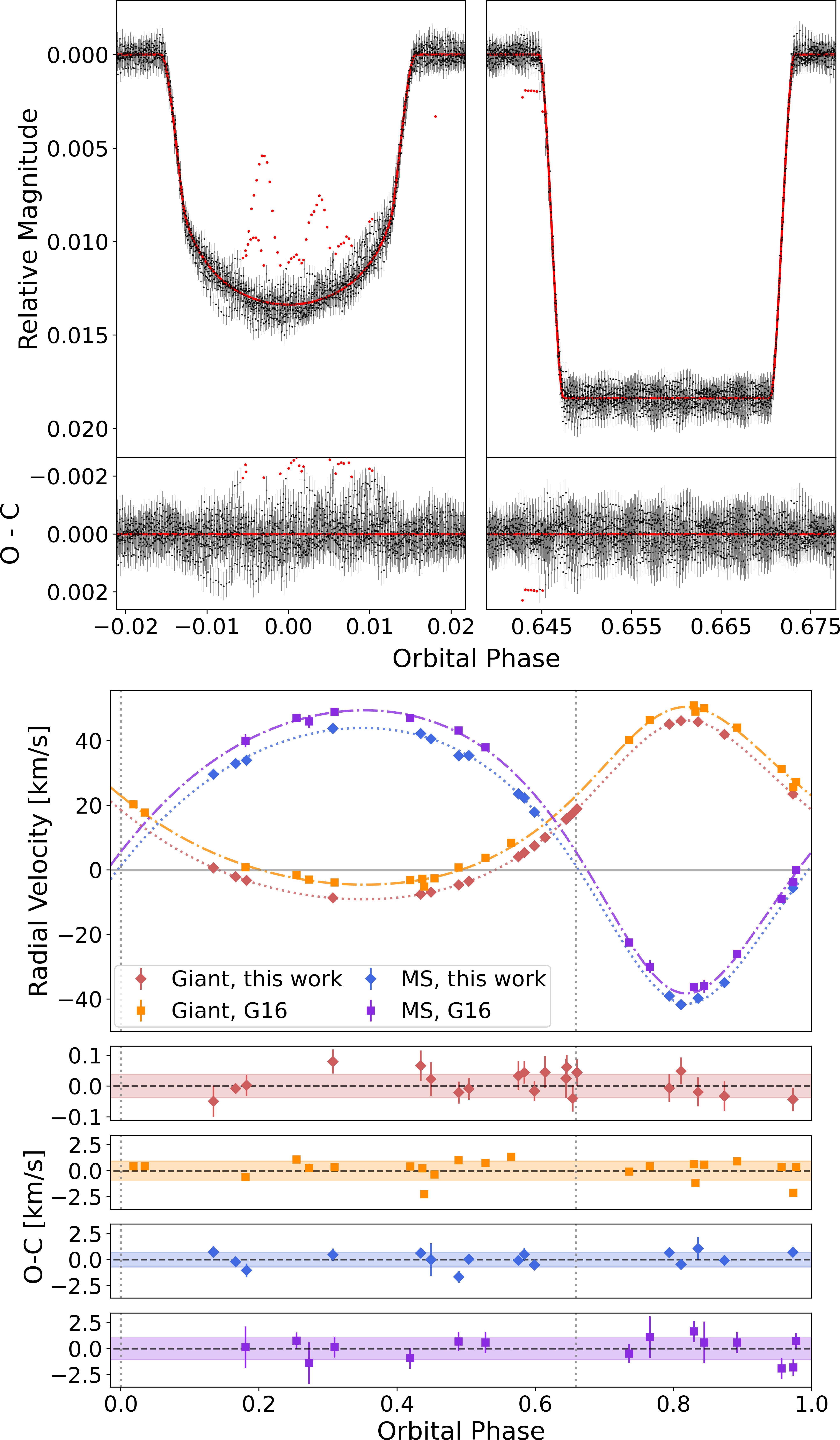}
    \caption{Binary Model fits to {\it Kepler} PDCSAP light curve (upper left and right panels) and radial velocities (lower 5 panels) for KIC\,8430105. Red markers in the light curve indicate sigma-clipped data. In RV panels: Red diamonds indicates our giant component RVs, while blue is for the MS component. Orange (giant) and purple (MS) squares represent measurements from \citetalias{Gaulme2016}. The models for \citetalias{Gaulme2016} RVs shown here have been fitted independently of that paper. Top O-C: Residuals for our RVs of the giant component. O-C shaded areas indicate corresponding unweighted standard deviations. Upper middle O-C: Residuals of the \citetalias{Gaulme2016} RVs of the giant component relative to the model in the RV plot. Lower middle O-C: Residuals for our RVs of the MS component. Bottom O-C: Residuals of the \citetalias{Gaulme2016} RVs of the MS component relative to the model in the RV plot. Grey vertical dotted lines at phase $0.0$ and $\sim0.66$ in RV panels indicate location of eclipses.}
             \label{fig:8430105}%
    \end{figure}

In Fig.~\ref{fig:8430105}, the best-fit JKTEBOP model is compared to the observed light curve and measured radial velocities. The upper-most panels show the normalized and truncated \emph{Kepler} PDCSAP light curve, with corresponding model in red. It is clear from the light curve O-C diagram just below that the residuals are dominated by a combination of solar-like oscillations and spot eclipsing during annular eclipse of the giant, rather than random errors. The JKTEBOP residual permutation method is supposed to estimate uncertainties in the presence of correlated noise by assuming that all residuals are correlated. However, it has been criticized by \citet{cubillos2017} for not corresponding to a sound resampling procedure. Its statistical properties were also not established by the author \citep{Southworth2008}. Therefore, we implemented a residual block bootstrap to estimate uncertainties. We also compared with estimates obtained using the residual permutation method and the JKTEBOP Monte Carlo simulation in Table~\ref{table:uncertainty} in the appendix. The JKTEBOP residual permutation method produced uncertainties that were relatively consistent with our residual block bootstrap when using our RVs from FIES, while the JKTEBOP Monte Carlo simulation produced significantly lower estimates. See Appendix~\ref{sec:uncertainty} for the full comparison between the methods.

\begin{figure}
   \centering
    \includegraphics[width=\columnwidth]{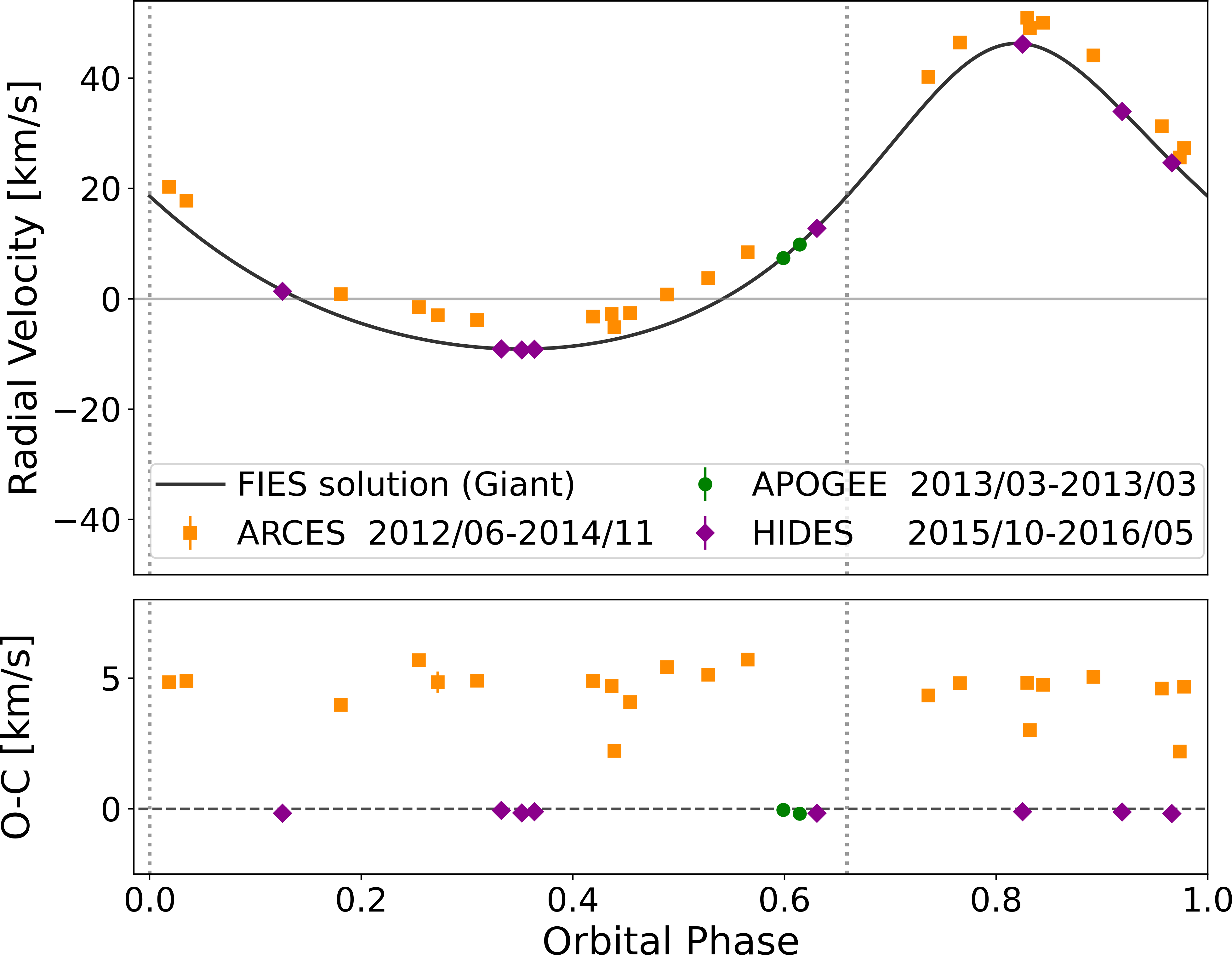}
    \caption{Comparison between our RV curve for the RG obtained with FIES RVs (black line), and the RVs of three different spectrographs: ARCES \citepalias{Gaulme2016} (orange squares), HIDES \citep{Helminiak2016} (purple diamonds), and APOGEE (green circles).}
             \label{fig:hides_sdss}%
\end{figure}

The 5 lower panels in Fig.~\ref{fig:8430105} show phased radial velocities and O-C diagrams. Included are our RVs as red and blue diamonds for the giant and MS star, respectively. Also included for comparison are \citetalias{Gaulme2016} RVs as orange (giant) and purple (MS) squares. The RV models are all fitted by us, including those used for the \citetalias{Gaulme2016} RVs. A large system velocity offset between $4.5$ and $4.7$ km/s is present between our RVs and those of \citetalias{Gaulme2016}. Patrick Gaulme (private communication) noticed that this offset is also present between the \citet{Gaulme2016} RVs and observations from two other spectrographs: \citet{Helminiak2016} published 8 RVs for the giant captured with the HIDES spectrograph at Okayama Astrophysical Observatory \citep{Izumiura1999}, and 2 observations are available from APOGEE \citep{APOGEE2017, APOGEEdr11}. In Fig.~\ref{fig:hides_sdss} these observations, and the ARCES RVs of the giant from \citetalias{Gaulme2016}, are compared with the solution obtained with FIES RVs. The velocity offset of the ARCES observations is clearly either instrumental or somehow from the templates used, and not astrophysical in nature.

The 4 RV O-C diagrams in Fig.~\ref{fig:8430105} show residuals for the giant and MS component RVs relative to the models. Root mean square (RMS) of the residuals of our RVs from FIES is $40$\,m/s for the giant and $0.7$\,km/s for the MS component. The high precision is primarily due to a combination of the following factors: A high resolution and S/N ratio for the observations, as well as a successful wavelength calibration involving per-observation Thorium-Argon comparison calibrations taken just before each observation.

Comparing, the HIDES RVs have a mean velocity offset from the FIES RV model for the giant of $-0.14$\,km/s in Fig.~\ref{fig:hides_sdss}. After correcting the HIDES measurements for the mean offset, we measure an RMS of $40$\,m/s for the residuals around our own model from FIES RVs. This is the same precision as for the FIES RVs themselves, and better than the RMS $71$\,m/s reported in \citet{Helminiak2016}. This indicates that the RV precision available in the HIDES RVs is $\leq40$\,m/s, and confirms that our model for the RV curve of the giant is precise to this level, up to a small instrumental velocity offset.

Properties of KIC\,8430105 determined by our eclipsing binary analysis are presented in Table~\ref{table:EBdata}.
\subsection{Activity and third light}
Since contamination in the light curve by third light reduces the relative eclipse depths, unmeasured third light can have significant effects on the found stellar radii and inclination. While the light curve might contain \emph{some} independent information on third light, it is generally not possible to fit for it freely from other parameters \citep{Southworth2010}. The \emph{Kepler} PDCSAP light curves are corrected, quarter-to-quarter, for a median contamination estimate based on the estimated point-spread-function of known, nearby stars \citep[section XIII]{PDCSAP_3_2017}. However, guarantees on the accuracy of these estimates are limited.

In Appendix~\ref{sec:tlight}, we qualitatively examine the target pixels files for possible contamination by nearby stars and quantitatively investigate the effect of third light on the parameters for KIC\,8430105. We find that, quite naturally, the relative accuracy that we require for the third light estimate depends significantly on the actual amount of third light present relative to the luminosity ratio of the system. To illustrate, the significance on found parameters was very limited for the \emph{Kepler} PDCSAP light curve. For this light curve, we could assume that the \emph{Kepler} pipeline has properly corrected for any contamination present and fix third light to 0.0 in JKTEBOP when measuring the best fit. When estimating uncertainties for the \emph{Kepler} PDCSAP light curve, we could then provide, as a prior, an external third light observation to JKTEBOP of $0.0$ with uncertainty of $\pm100\%$ the maximum \emph{Kepler} pipeline estimate of $0.005$, in order to account for the possibility that the mean third light estimate is different from the true mean third light. This was not the case for the TESS light curve, where the reported contamination estimate for sector 15 was $0.076$, multiple times larger than the light fraction of the MS component. We also find reason to suspect that the KASOC \emph{Kepler} light curve might contain significantly more third light than the mean \emph{Kepler} pipeline estimate. In Appendix~\ref{sec:uncertainty} we find that large amounts of third light becomes a problem for both the best fit evaluation and the uncertainty estimation when inclination is close to $90^\circ$ and a strict prior is not provided. Based on these results we elect to assign weight only to measurements produced using the \emph{Kepler} PDCSAP light curve. When we reference our processed \emph{Kepler} light curve during later analysis, we refer only to the PDCSAP light curve, and not the KASOC light curve unless clearly specified.

Magnetic activity can also affect the eclipse depths. The red giant in KIC\,8430105 displays significant magnetic variability in its light curve, larger than the eclipse depth for about half of the 4 years covered by \emph{Kepler}. \citet{Gaulme2014} measured the stellar variation period of the giant to be very close to 2:1 synchronized to the orbital period, and the maximum peak-to-peak amplitude to be $7\%$. \citet{Benbakoura2021} also measured the photometric magnetic activity level proxy to be $S_{\text{ph}} = 1.2\%$.

To estimate the effect that the activity can have on the found parameters, we compared solutions obtained using either eclipses obtained near flux maxima or minima. We selected 10 eclipses for each component, 5 around maximum and 5 around minimum. Then, we ran new models with either both components at one of the extrema, or one at maximum and one at minimum. We fixed the orbital period, reference $t_G$, limb darkening parameters, and the light scale factor to the original best fit. We also fixed the inclination, since two of the combinations became locked to $i=90^\circ$ otherwise. The radius of the giant is well within the uncertainties in all cases, changing with only up to $0.16\sigma$. The main sequence radius is only significantly changed, larger by $2\sigma$, in the presence of flux minima in the annular eclipse where spot eclipsing also occurs.
\subsection{Comparing with dynamical results from \citetalias{Gaulme2016}}
Our dynamical mass and radius measurements in Table~\ref{table:EBdata} are significantly lower than those reported by \citetalias{Gaulme2016} for both stellar components, with all results different by more than the one-sigma combined uncertainty limits of the reported values. Specifically, using their reported uncertainties, our mass of the giant is lower by $2.8\sigma$, and our radius by $3.5\sigma$. Our mass of the MS star is lower by $1.7\sigma$, and the radius of the MS star by $4.1\sigma$.

We attempted to reproduce the \citetalias{Gaulme2016} findings using our normalized \emph{Kepler} PDCSAP light curve, no sigma-clipping, and the \citetalias{Gaulme2016} RVs with their reported RV errors. For this we used their spectroscopic temperature of the giant of 5042K, their spectroscopic metallicity $\rm [Fe/H] = -0.49$, and our microturbulence of $\rm 0.91 km s^{-1}$ for limb darkening parameter interpolation. All of their results were recovered for masses, radii and inclination within $1\sigma$. However, the fit had a reduced $\chi^2$ value of 1820 for the RVs of the giant, indicating that the reported RV uncertainties are severely underestimated. 

Adjusting these to obtain $\chi^2 \sim 1$, by adding a common RV uncertainty of $0.88$\,km/s to all RVs of their giant component, resulted in an increase in inclination of $3$ times their reported $1\sigma$ uncertainty, as the light curve was weighted higher during fitting. The radius of the MS star was also $1.5\sigma$ lower than reported by \citetalias{Gaulme2016}.

We tried using $4\sigma$ clipping of the light curve along with the adjusted RV uncertainties for their RVs of the giant. This increased the inclination, slightly lowered the radii, and left the masses unchanged. While the radii were lowered, this was nowhere near enough to bridge the gap to our results reported in Table~\ref{table:EBdata} using our own spectroscopic follow-up RVs. Our radii found using their RVs were still larger than ours by $\sim2.5$ times the \citetalias{Gaulme2016} reported uncertainty for both the giant and MS star.

When performing uncertainty estimation using the \citetalias{Gaulme2016} RVs with a common addition to RV uncertainty for the giant, we obtained a global increase in the uncertainty estimates of all parameters, suggesting that the uncertainties of \citetalias{Gaulme2016} for the stellar parameters are underestimated.

From the above investigation we are confident in concluding that while differences between the processed \emph{Kepler} light curves of \citetalias{Gaulme2016} and this study did have some effect on the parameters found, the primary difference by far stems from the separate spectroscopic follow-up RV observations. Paper I noted that the RVs in \citetalias{Gaulme2016} seem to suffer from large epoch-to-epoch drifts. We find indications of this for KIC\,8430105 as their RV uncertainties for the giant component are much smaller than the residuals in a binary solution. We suspect that this is caused by drifts in the wavelength calibration between observations. \citet{Rawls2016} comments that some of their early ARCES observations "had insufficiently frequent ThAr calibration images to arrive at a reliable wavelength solution", which might be the cause, given that \citetalias{Gaulme2016} also use ARCES spectra. In \citet[Table C.1]{Benbakoura2021}, and in \citetalias{Gaulme2016}  (Patrick Gaulme, private communication), all RV uncertainties for ARCES data were replaced with $0.5$\,km/s for the giant and $1.0$\,km/s for their analysis.
\begin{table*}
\centering
\caption{\label{table:EBdata}Properties of the eclipsing binary KIC\,8430105.}
\begin{threeparttable}
\begin{tabular}{lcc}
\hline
\hline
Quantity & \emph{Kepler} & TESS\tnote{a} \\
\hline
RA  (J2000)\tnote{b}                                & $19:26:14.069$   & $19:26:14.07$   \\
DEC (J2000)\tnote{b}                                & $+44:29:17.48$   & $+44:29:17.4$   \\
Magnitude ($K_p$, $T$)		                        &    10.420        & 9.813    \\ 
\hline
$T_{\rm eff,RG}$ (K)                                &\multicolumn{2}{c}{4990 (80)}      \\
$T_{\rm eff,MS}$ (K) \tnote{c}                                   & $5655 (80)$     & $5706 (80)$    \\
$[\rm Fe / \rm H]$ (dex)                                        &\multicolumn{2}{c}{-0.46 (10)}\\
$[\rm \alpha / \rm Fe]$ (dex)                            &\multicolumn{2}{c}{0.08 (10)}\\
\hline
Orbital period (days)                               & $63.327106 ^{+64}_{-65}$  & fixed   \\[0.1cm]
Reference time $\rm t_{\rm RG}$ (days)                     & $54998.2336^{+16}_{-15}$ & $54998.2342_{-37}^{37}$  \\[0.1cm]
Inclination $i$ ($\circ$)                           & $89.56^{+32}_{-18}$     &   $89.60_{-64}^{+35}$    \\[0.1cm]
Eccentricity $e$                         & $0.25644^{+10}_{-10}$     & $0.25643_{-14}^{+14}$     \\[0.1cm]
Periastron longitude $\omega$ ($\circ$)             & $349.41^{+11}_{-12}$     & $349.49_{-12}^{+11}$      \\[0.1cm]
$e\cos\omega$                                       & $0.252100^{+40}_{-40}$       & $0.25213_{-11}^{+11}$ \\[0.1cm]
$e\sin\omega$                                       & $-0.04699^{+52}_{-54}$      & $-0.04677_{-56}^{+50}$ \\[0.1cm]
Sum of the fractional radii $r_{\rm MS}+r_{\rm RG}$ & $0.09656^{+32}_{-27}$    & $0.09621_{-43}^{+13}$   \\[0.1cm]
Ratio of the radii $k$                              & $0.10025^{+28}_{-27}$    & $0.0981_{-17}^{+17}$   \\[0.1cm]
Surface brightness ratio $J$                        & $1.6499^{+79}_{-80}$      & $1.538_{-18}^{+18}$     \\[0.1cm]
$\frac{L_{\rm MS}}{L_{\rm RG}} (K_p, T)$            & $0.017087^{+53}_{-54}$    & $0.01555_{-53}^{+55}$  \\[0.1cm]
$K_{\rm RG}$  (km/s)                                      & $27.697^{+14}_{-12}$     & $27.694_{-12}^{+13}$    \\[0.1cm]
$K_{\rm MS}$  (km/s)                                      & $42.70^{+22}_{-22}$      & $42.70_{-22}^{+22}$      \\[0.1cm]
semi-major axis $a (R_{\odot})$                 & $85.18^{+26}_{-26}$     & $85.17_{-27}^{+27}$      \\[0.1cm]
$\gamma_{\rm RG}$   (km/s)                                & $11.615^{+10}_{-9}$    & $11.6199_{-79}^{+98}$     \\[0.1cm]
$\gamma_{\rm MS}$   (km/s)                               & $12.03^{+16}_{-16}$     & $12.03_{-16}^{+16}$     \\[0.1cm]
Mass$_{\rm RG}(M_{\odot})$                          & $1.254^{+14}_{-14}$      & $1.254_{-14}^{+14}$      \\[0.1cm]
Mass$_{\rm MS}(M_{\odot})$                          & $0.8134^{+50}_{-50}$      & $0.8133_{-51}^{+51}$      \\[0.1cm]
Radius$_{\rm RG}(R_{\odot})$                        & $7.475^{+32}_{-30}$     & $7.467_{-42}^{+100}$     \\[0.1cm]
Radius$_{\rm MS}(R_{\odot})$                        & $0.7494^{+42}_{-39}$      & $0.734_{-15}^{+16}$      \\[0.1cm]
log$g_{\rm RG}$ (cgs)                               & $2.7892^{+32}_{-34}$     & $2.7902_{-114}^{+44}$       \\[0.1cm]
log$g_{\rm MS}$ (cgs)                               & $4.5990^{+37}_{-41}$     & $4.616_{-18}^{+17}$       \\[0.1cm]
$\rho_{\rm RG} (10^{-3}\rho_{\bigodot})$            & $3.003^{+25}_{-30}$     & $3.018_{-121}^{+36}$                   \\[0.1cm]
$\rho_{\rm MS} (\rho_{\bigodot})$                   & $1.933^{+25}_{-28}$     & $2.05_{-13}^{+13}$                  \\[0.1cm]
Third light $L_3$ ($\%$)                                  & $0.00^{+11}_{-12}$ \tnote{d}         & $6.0_{-31}^{+31}$ \tnote{e} \\[0.1cm]
\hline
ld$_{a, \rm RG}$                                     & $0.489^{+13}_{-13}$      & $0.427_{-36}^{+35}$\\[0.1cm]
ld$_{a, \rm MS}$                                    &  $0.3758867$ \tnote{f}      & $0.2780412$ \tnote{g}\\
ld$_{b, \rm RG}$                                     & $0.1978336$ \tnote{f}      & $0.2360113$ \tnote{g}\\
ld$_{b, \rm MS}$                                    &  $0.2767570$ \tnote{f}      & $0.2845344$ \tnote{g}\\
$\frac{L_{\rm MS}}{L_{\rm RG}}$ bolometric          & $0.01658$      & $0.01654$       \\
\hline
Gaia distance (pc)\tnote{h}                        &\multicolumn{2}{c}{$726.3 (66)$} \\
Magnitude $m_G$\tnote{h}                               &\multicolumn{2}{c}{$10.400$} \\
$A_{ G}$\tnote{h}                               &\multicolumn{2}{c}{$0.139 (16)$} \\
$BC_{ G}$\tnote{h}                              &\multicolumn{2}{c}{$-0.034 (23)$} \\
Gaia Radius$_{\rm RG} (R_{\odot})$ \tnote{h}                 &\multicolumn{2}{c}{$7.71(28)$}\\
Dynamical age estimate (Gyr)\tnote{i}                        & \multicolumn{2}{c}{$3.7 (4)$} \\ 
\hline
\end{tabular}

\begin{tablenotes}
		\scriptsize
\item[a] TESS dynamical values are from the median solution from our residal block bootstrap and not the best fit to the original TESS light curve. See Appendix~\ref{sec:tlight}~and~\ref{sec:uncertainty}.
\item[b] From the KIC and TIC. 
\item[c] From dynamical analysis.
\item[d] Fixed to 0.0 for best fit. Constrained to $0.0\%\pm 0.5\%$ during uncertainty estimation. See Appendix~\ref{sec:uncertainty}.
\item[e] Constrained to $7.6\%\pm 4.0\%$. See Appendix~\ref{sec:uncertainty}.
\item[f] Interpolated from \citet{Claret2011}.
\item[g] Interpolated from \citet{claret2017}.
\item[h] See Section~\ref{sec:distance}.
\item[i] See Section~\ref{sec:age}.

\end{tablenotes}
\end{threeparttable}
\end{table*}

\begin{table*}
\centering
\caption{\label{table:RGdata}Asteroseismic and dynamical measurements of the RG in KIC\,8430105. }
\begin{threeparttable}
\begin{tabular}{lcc}
\hline
\hline
Quantity & \multicolumn{2}{c}{Value} \\
\hline
$\Delta\nu_{\rm ps}$ ($\mu$Hz)              &\multicolumn{2}{c}{$7.04 (4)$}\\
$\Delta\nu_0$ ($\mu$Hz)                                     & \multicolumn{2}{c}{$7.123 (35)$}\\
$\Delta \nu_{\rm c}$ ($\mu$Hz)                        &\multicolumn{2}{c}{$7.081(41)$}\\
$\delta \nu_{\rm 0, 2}$ ($\mu$Hz)                        &\multicolumn{2}{c}{$0.90(20)$}\\
$\nu_{\rm max}$ ($\mu$Hz)               &\multicolumn{2}{c}{$76.78 (81)$}\\
$\Delta P_{\rm obs}$ (s)                        &\multicolumn{2}{c}{$111.7(37)$}\\
$\epsilon$ ($\mu$Hz)                        &\multicolumn{2}{c}{$1.115(47)$}\\
$\epsilon_{\rm c}$ ($\mu$Hz)                        &\multicolumn{2}{c}{$1.164 (53)$}\\
$f_{\Delta\nu}$ correction factor\tnote{2}         &\multicolumn{2}{c}{$0.974(3)$}\\
$f_{\nu_{\rm max}}$ correction factor\tnote{3}     &\multicolumn{2}{c}{$1.015(5)$}\\
\hline
Mass$_{\rm dyn} (M_{\odot})$                        &\multicolumn{2}{c}{$1.254(14)$}\\
Mass$_{\rm seis-raw} (M_{\odot})$                  &\multicolumn{2}{c}{$1.586(70)$}\\
Mass$_{\rm seis-corr, a} (M_{\odot})$\tnote{1, 2}                  &\multicolumn{2}{c}{$1.428(66)$}\\
Mass$_{\rm seis-corr, b} (M_{\odot})$\tnote{1, 3}                  &\multicolumn{2}{c}{$1.517(71)$}\\
Mass$_{\rm seis-corr, c} (M_{\odot})$\tnote{1, 2, 3}                  &\multicolumn{2}{c}{$1.365(66)$}\\
Mass$_{\rm seis-mode} (M_{\odot})$\tnote{4}                  &\multicolumn{2}{c}{$1.23(6)$}\\
\hline
Radius$_{\rm dyn} (R_{\odot})$                  &\multicolumn{2}{c}{$7.475(31)$}\\
Radius$_{\rm seis-raw} (R_{\odot})$                  &\multicolumn{2}{c}{$8.29(14)$}\\
Radius$_{\rm seis-corr, a} (R_{\odot})$\tnote{1, 2}                  &\multicolumn{2}{c}{$7.86(14)$}\\
Radius$_{\rm seis-corr, b} (R_{\odot})$\tnote{1, 3}                  &\multicolumn{2}{c}{$8.16(14)$}\\
Radius$_{\rm seis-corr, c} (R_{\odot})$\tnote{1, 2, 3}                  &\multicolumn{2}{c}{$7.74(14)$}\\
Radius$_{\rm seis-mode} (M_{\odot})$\tnote{4}                  &\multicolumn{2}{c}{$7.45(11)$}\\
\hline
$\rho_{\rm dyn} (10^{-3}\rho_{\odot})$                &\multicolumn{2}{c}{$3.003(26)$}\\
$\rho_{\rm seis-raw} (10^{-3}\rho_{\odot})$                &\multicolumn{2}{c}{$2.788(27)$}\\
$\rho_{\rm seis-corr, a} (10^{-3}\rho_{\odot})$\tnote{1, 2}                &\multicolumn{2}{c}{$2.939(34)$}\\
\hline
$\log g_{\rm dyn} (\rm dex)$                   &\multicolumn{2}{c}{$2.7892(33)$}\\
$\log g_{\rm seis-raw} (\rm dex)$                   &\multicolumn{2}{c}{$2.8017(58)$}\\
$\log g_{\rm seis-corr, b} (\rm dex)$ \tnote{1, 3}                  &\multicolumn{2}{c}{$2.7953(61)$}\\
\hline
\end{tabular}
\begin{tablenotes}
		\scriptsize
\item[1]Asteroseismic scaling relations using corrections are marked \emph{seis-corr,x} with x=a for \citet{Rodrigues2017} $\Delta\nu$ corrections, x=b for \citet{Viani2017} $\nu_{\rm max}$ corrections, and x=c for both. Uncorrected scaling relations are marked \emph{seis-raw}.
\item[2] Correction to $\Delta\nu$ scaling relation according to \citet[Figure 3]{Rodrigues2017} assuming RGB star with [Fe/H] = -0.5 and $M = 1.3M_{\bigodot}$. Uncertainty estimated based on track resolution.
\item[3] Correction to $\nu_{\rm max}$ scaling relation according to \citet[Figure 3]{Viani2017} assuming $T_{\rm eff} = 4990K$ or $\log g = 2.8$ and [Fe/H]$=-0.5$. Uncertainty estimated from varying atmosphere model or $T_{\rm eff}$.
\item[4] Initial forward modelling of the individual mode frequencies and metallicity by \citet{buldgen2019}.
\end{tablenotes}
\end{threeparttable}
\end{table*}
%
%

\section{Gaia distance determination and comparison}\label{sec:distance}
With parallax and apparent magnitude provided by the Early Third Gaia Data Release \citep{gaia2021} an independent distance estimate to KIC\,8430105 is available. With a raw parallax of $1.358 \pm 0.012$ mas and correction of the parallax zero-point of $-0.0186$ mas determined using the python software \emph{gaiadr3-zeropoint} \citep{lindegren2021}, the distance is $726.3 \pm 6.6$pc. The apparant magnitude of KIC\,8430105 in the Gaia \emph{G}-band is $m_G = 10.400$.

To account for extinction, we used the online tool by \citet{lallement2019} to obtain an extinction at 5500Å, $A_0 = 0.167$. This was then converted to $A_G = 0.1385$ using a Gaia EDR3 Extinction Law that is part of the public auxiliary data provided by ESA/Gaia/DPAC/CU5 and prepared by Carine Babusiaux\footnote{\url{https://www.cosmos.esa.int/web/gaia/edr3-extinction-law}}. To obtain an uncertainty estimate on $A_G$, we also derived this parameter using the \emph{Bayestar19} dustmap \citep{green2019, green2018}. Here, we converted to $E(B-V)$ following the method outlined on their web-page\footnote{\url{http://argonaut.skymaps.info/usage}}, obtaining two separate results. Then, we used \citet[Table 2]{casagrande2018b} to calculate an extinction coefficient for the RG in the Gaia $G$ band, extrapolating to slightly outside of the temperature range from which the fit was deemed valid. From this we obtain two separate extinction estimates, $A_{G, *} = (0.1371, ~ 0.1544)$. We adopt the largest deviation from the original extinction estimate as the uncertainty. Therefore, we end up with $A_G = 0.139 \pm 0.016$.

A bolometric correction $BC_G = -0.034\pm 0.023$ was calculated using the interpolation routines of \citet{casagrande2018a} for Gaia photometry \citep{casagrande2018b}. The uncertainty on the bolometric correction was estimated by taking the largest variation induced by either changing the metallicity 0.1dex or the effective temperature by 80K. 

The Gaia estimate for the luminosity of the RG was then determined using the dynamically derived bolometric light ratio:
\begin{align}
    m_{\rm RG} &= -2.5 \log_{10} \left(\frac{1}{1+ L_B/L_A}\right) + m_G, \\
    \mathcal{M}_{\rm bol} &= m_{\rm RG} + 5 - 5\log_{10}(\frac{d}{pc}) - A_G + BC_G, \\
    L_{\rm RG, Gaia} &= 10^{0.4(\mathcal{M}_\odot - \mathcal{M}_{\rm bol})} = 33.1 \pm 1.0 L_\odot.
\end{align}

We then estimated the radius of the RG from the Gaia luminosity and our spectroscopic temperature to be
\begin{align}
    R_{\rm RG, Gaia} = 7.71 \pm 0.28 R_{\odot}.
\end{align}
Overall, this is in agreement with, but less precise than, the dynamically determined radius of $7.475 \pm 0.031 R_\odot$.
\section{Dynamical age estimate}\label{sec:age}
A model-dependent age of the components in the system was estimated by comparing our dynamical measurements to PARSEC isochrones from \citet{bressan2012}, obtained through the CMD 3.6 web interface\footnote{\url{http://stev.oapd.inaf.it/cgi-bin/cmd}}. Since the RG has a small $\alpha$-element enhancement of $[\alpha /\rm Fe] = 0.08\pm 0.04$, the metallicity must be scaled if we use isochrones with no enhancement. In the current literature, only \citet{Salaris1993} has investigated whether such a scaling is possible. They investigated low mass ($0.6-1.0 M_{\odot}$), very low metallicity ($Z=10^{-4} - 10^{-3}$), and high $\alpha$ enhancement ($[\alpha / \rm Fe] = 0.6-0.9$) for use with globular clusters. This is not the regime that our RG is in. For a comparison of the \citet{Salaris1993} correction with a case slightly closer to ours, see \citet[Figure 1]{Miglio2021} which compares an 11Gyr, $[\rm Fe/H] = -0.62$, $[\alpha / \rm Fe] = 0.2$ isochrone with one using a corrected metallicity of $[\rm Fe/H] = -0.5$ and no $\alpha$ enhancement.
In this work we assume that using the metallicity correction by \citet{Salaris1993},
\begin{align}
    [\rm M/H] \simeq [\rm Fe/H] + \log_{10} \left(0.638 \cdot 10^{[\alpha/\rm Fe]} + 0.362 \right),
\end{align}
is better than using no correction at all. With this we obtain a corrected $[\rm M/H] = -0.41 \pm 0.10$ for KIC\,8430105.

\begin{figure}
   \centering
   \includegraphics[width=8.6cm]{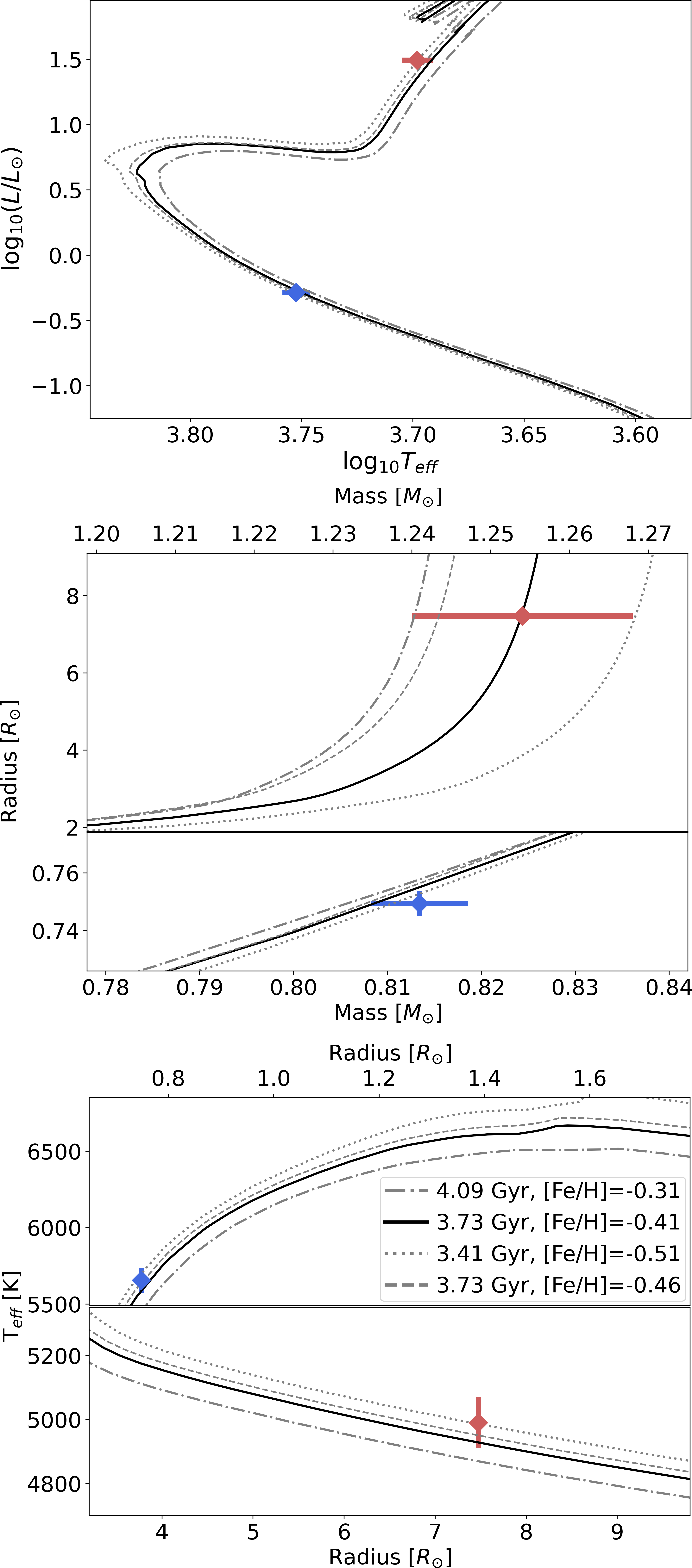}
   \caption{HR, Mass--Radius and radius--$T_{\rm eff}$ diagrams with measurements of the eclipsing binary components compared to representative isochrones of different ages and metallicities as indicated by the legend in the bottom panel.}
             \label{fig:mrt}
\end{figure}

We matched isochrones by eye from a grid of PARSEC isochrones with ages spaced 5Myr apart. Fig.~\ref{fig:mrt} shows HR, Mass--Radius and radius--$T_{\rm eff}$ diagrams with our measurements from the eclipsing binary and spectroscopic analysis. Overlaid are representative isochrones of different ages and metallicities. Included is an isochrone assuming no enhancement of $\alpha$-elements ($[\rm Fe/H] = -0.46$) and the same age as the metallicity-corrected best-matching isochrone ($[\rm Fe/H] = -0.41]$). An age estimate for the system is found by matching isochrones to the dynamical mass and radius of the giant in the mass/radius diagram, in a similar manner as Paper I. As also illustrated in Paper I, uncertainty in age is primarily due to uncertainty in mass and metallicity; the radius provides a negligible contribution as the isochrones are close to vertical near the RG component. We obtained 4 different ages when simultaneously varying metallicity of the isochrone by $\pm0.1$dex and dynamical mass by $\pm1\sigma$. We estimated the $1\sigma$ age uncertainty as the largest deviation in age brought on by this. With this, we have estimated the age of the system to be $3.7\pm 0.4$ Gyr.
\section{Comparing with asteroseismology}
\label{sec:compare}
We have obtained accurate dynamical measurements of mass and radius, and a spectroscopic determination of $T_{\rm eff}$ for the RG. It is now possible to compare with asteroseismic measurements in order to ascertain if this system deviates significantly even after $\Delta\nu$ corrections are applied, as was found by \citetalias{Gaulme2016}. 

The global asteroseismic parameters, extraction detailed during Sect.~\ref{sec:seismology}, are used with the scaling relations Eq.~\ref{eq:01}--\ref{eq:04} to derive stellar mass, radius, density and surface gravity for the RG. A metallicity dependent model correction to the $\Delta\nu$ scaling relation for average density from \citet{Rodrigues2017} is examined, along with a temperature-dependent model correction to the $\nu_{\rm max}$ scaling relation for $\log g$ by \citet{Viani2017}. 
Measurements are detailed in Table~\ref{table:RGdata}, and Mass--Radius plots are shown along with 1-D plots of average density $\rho$ and surface gravity $\log g$ in Fig.~\ref{fig:mr_ast}. Dynamical measurements are also compared with individual frequency modelling by \citet{buldgen2019} and \citet{Joergensen2020}.
\subsection{Asteroseismology of KIC\,8430105}\label{sec:seismology}
The power density spectrum was calculated by unweighted least-squares sine-wave fitting to the KASOC filtered \emph{Kepler} light curve \citep{Frandsen1995}. We determined $\nu_{\rm max}$ using multiple different methods, obtaining several different results, which is outlined in Sect.~\ref{sec:numax}. The result in Table~\ref{table:RGdata} was found using least-squares fitting of a Gaussian power excess together with a background model \citet[Eq. 5]{Handberg2017}, as in \citet{Arentoft17}.
%
%

To determine $\Delta\nu_{\rm ps}$, we used a method described in \citet{Brogaard21, Arentoft19}, where the power spectrum is split into bins of trial $\Delta\nu$ and stacked. The largest peak is then obtained when the correct $\Delta\nu$ is used. We identified individual oscillation modes using the method described in \citet{Arentoft17}. From these, six detected radial modes where used to obtain a refined large frequency spacing $\Delta\nu_0$. The observed period spacing of the $\ell = 1$ modes, $\Delta P_{\rm obs}$, was also measured.

The measured and adopted asteroseismic parameter values are available in Table~\ref{table:RGdata}.
\subsubsection{$\nu_{\rm max}$ determination}\label{sec:numax}
To determine $\nu_{\rm max}$ we first performed a least-squares fit to the power density spectrum as in \citet{Arentoft17} using a Gaussian power excess model with a background profile following \citet[Eq. 5]{Handberg2017} (hereafter H5). This yielded $\nu_{\rm max} = 76.78 \pm 0.81 \mu$Hz, which is in very good agreement with $\nu_{\rm max} = 76.70 \pm 0.57 \mu$Hz as determined by \citetalias{Gaulme2016} using the Bayesian DIAMONDS pipeline \citet{Corsaro2014} following the methodology explained by \citet{Corsaro2015}. \citetalias{Gaulme2016} also state that they obtained a fully consistent result using a Bayesian maximum a posteriori method \citep{Gaulme2009}, though they did not give numbers.

However, the power density spectrum of KIC\,8430105 has low S/N due to magnetic suppression \citep{Gaulme2014, Benbakoura2021} and a smoothed version of the power spectrum is slightly asymmetric. As in other cases with magnetic suppression \citep{Arentoft17} we found that using our least-squares fit with a \citet[Eq. 4]{Handberg2017} (H4) background yielded a significantly different value, $\nu_{\rm max} = 73.03 \pm 0.86 \mu$Hz. From intercomparison of asteroseismic masses among clump stars in NGC\,6811 with $\nu_{\rm max}$ determined using Eq. (H5) and (H4), we know that this latter result is less trustworthy, but significant differences between results using a least-squares fit with (H5) and (H4) seems to be present only in cases with magnetic suppression. Specifically, we have not encountered this for any cases where S/N is low for observational reasons, e.q. stars of fainter magnitude in NGC\,6791 \citep{Brogaard21}.

Due to the potential effects on $\nu_{\rm max}$ due to magnetic suppression, we also decided to determine $\nu_{\rm max}$ with an alternative method and multiple background formulations. Using the procedure of \citet{Lund2017} involving affine-invariant Markov chain Monte Carlo (MCMC) sampling with \emph{emcee} \citep{ForemanMackey2013}, we obtained four relatively higher, but internally consistent, values for $\nu_{\rm max}$ of $78.70${\raisebox{0.5ex}{\tiny$\substack{+1.14 \\ -1.24}$}} $\mu$Hz, $78.49${\raisebox{0.5ex}{\tiny$\substack{+0.87 \\ -0.92}$}}, $78.49${\raisebox{0.5ex}{\tiny$\substack{+0.79 \\ -0.83}$}} $\mu$Hz, and $78.92${\raisebox{0.5ex}{\tiny$\substack{+1.02 \\ -1.09}$}} $\mu$Hz. These were obtained by using either the background model from \citet{Lund2017} described as a sum of two Harvey profiles with free exponents, timescales and amplitudes (first case), the H5 model (second case), the H4 model (third case), or a sum of two super-Lorentzian components as in \citetalias{Gaulme2016} (fourth case).

We note that a relatively significant difference exists between our least-squares result with (H5) and our MCMC results. While it is only around $1.5\sigma$, the values were not determined using independent data. The smoothed version of the background subtracted power density spectrum is asymmetric, likely resulting from two strong $l=0$ ($n=7,8$) modes around $58$ and $65 \mu$Hz, which makes a Gaussian fit of the power excess difficult and method dependent.

It is out of the scope of this paper to investigate in detail the deviation between these and other methods used to determine $\nu_{\rm max}$. For our continued analysis in this paper we have chosen to adopt the result of least-squares fitting with the (H5) background, $\nu_{\rm max} = 76.78 \pm 0.81 \mu$Hz, through a combination of the following considerations: (1) It has previously been found to produce consistent results for both high and low S/N power spectra of red giants. (2) It is very consistent with the independent analysis of the \emph{Kepler} light curve by \citetalias{Gaulme2016}, who found $\nu_{\rm max} = 76.70 \pm 0.57 \mu$Hz. (3) The difference between mass and radius found in Sect.~6.2, when comparing scaling relations and dynamical measurements, would only increase if we had instead adopted any of the MCMC results, thereby not changing the overall conclusions of this paper appreciably.
\subsection{Mass and radius}\label{sec:mr_ast}
\begin{figure}
    \centering
    \includegraphics[width=\columnwidth]{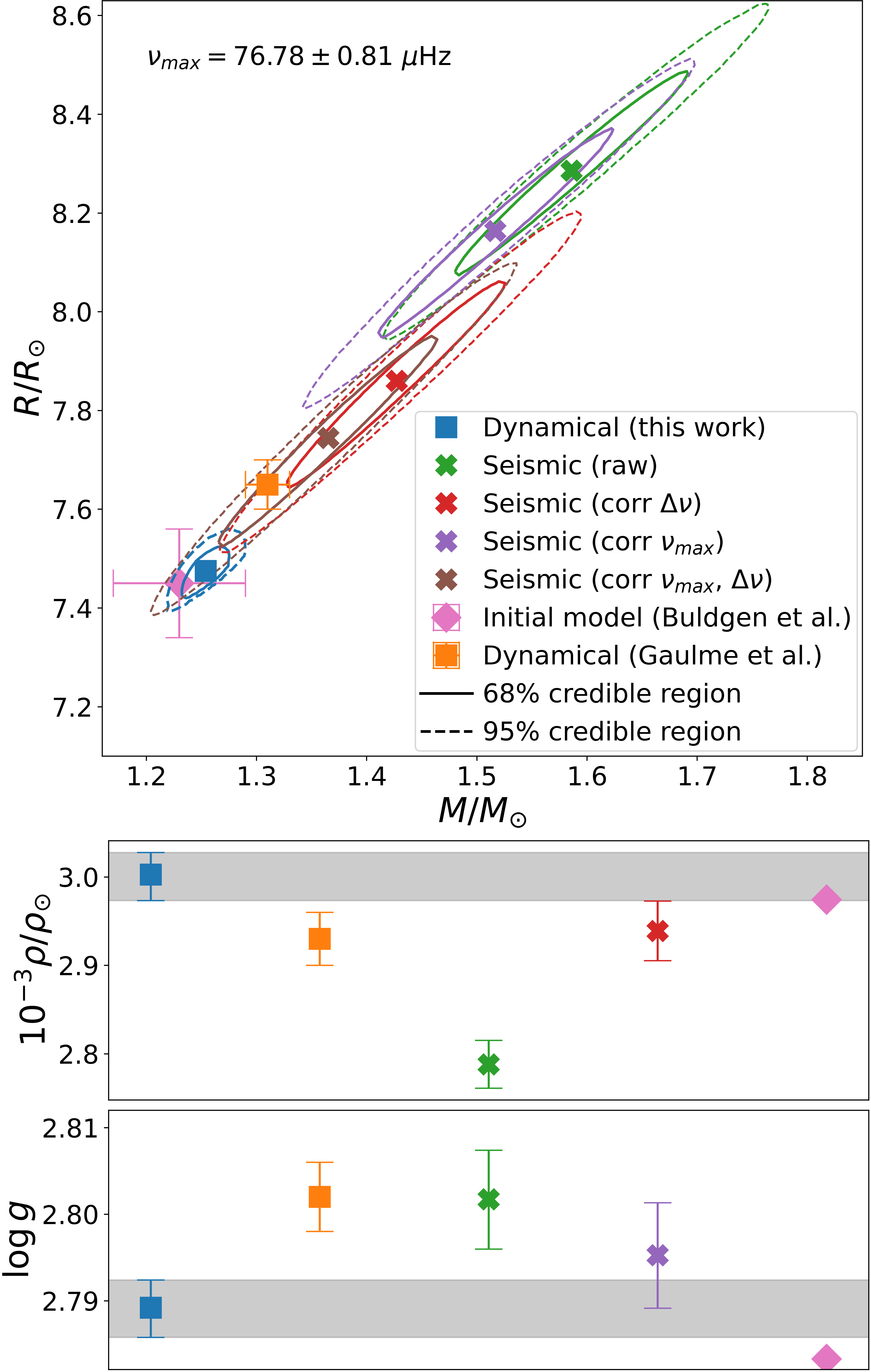}
    \caption{Upper panel: Mass--Radius plot for the giant with the results from Table~\ref{table:RGdata}. Solid contours indicate $1\sigma$ uncertainty (0.16, 0.84 quantiles), while dashed contours indicate $2\sigma$. Lower two panels: 1-dimensional density and surface gravity plots of the same results. Grey outlines indicate the dynamical $1\sigma$ limits from this work. Markers and colors in the 4 panels otherwise indicate: Our EB results (blue squares), EB results of \citetalias{Gaulme2016} (orange squares), uncorrected scaling relations (green crosses), scaling relations with $\Delta\nu$ correction from \citet{Rodrigues2017} (red crosses), scaling relations with $\nu_{\rm max}$ correction from \citet{Viani2017} (purple crosses), scaling relations with both corrections (brown crosses), and initial model of frequencies from \citet{buldgen2019} (pink diamonds). Uncertainty has not been estimated for the density and surface gravity from the \citet{buldgen2019} model.}
    \label{fig:mr_ast}
\end{figure}

In the upper panel of Fig.~\ref{fig:mr_ast} we show a Mass--Radius plot for the measurements included in Table~\ref{table:RGdata}. Included are our dynamical measurements using the \emph{Kepler} PDCSAP light curve and NOT RVs (blue squares), dynamical measurements from \citetalias{Gaulme2016} (orange squares), measurements using asteroseismic scaling relations without corrections (green), scaling relations with \citet{Rodrigues2017} correction to $\Delta\nu$ (red), with \citet{Viani2017} correction to $\nu_{\rm max}$ (purple), both corrections (brown) and an initial model from \citet{buldgen2019} using the individual mode frequencies along with the metallicity (pink diamonds).
The scaling relation results are all marked with crosses.
Two plotted contours for each of our dynamical and asteroseismic measurements indicate $1\sigma$ (solid) and $2\sigma$ (thin dashed) uncertainty for the respective results. The dynamical contours were calculated with a Gaussian kernel density estimation (KDE) of the residual block bootstrap results. The asteroseismic contours were calculated by the following steps: First, it was assumed that the uncertainties on $T_{\rm eff}$, $\Delta\nu_0$, $\nu_{\rm max}$, $f_{\Delta\nu}$, and $f_{\nu_{\rm max}}$ represent Gaussian $1\sigma$ standard deviations. Then, Monte Carlo sampling was done in order to propagate uncertainties to mass and radius. Finally, a Gaussian KDE was used to estimate the probability density functions and draw the contours.

The asteroseismic mass and radius from the uncorrected scaling relations is not consistent with our dynamical results.
This finding is in good agreement with that of \citetalias{Gaulme2016} and \citet{Brogaard2018}, who also finds that the uncorrected scaling relations overestimate mass and radius significantly for RGs. Mass is overestimated by $26\%$ relative to our dynamical result, while radius is overestimated by $11\%$.

The $\nu_{\rm max}$ correction from \citet{Viani2017} provides an insignificant improvement to mass and radius over the uncorrected scaling relation. Using the $\Delta\nu$ correction from \citet{Rodrigues2017}, we are able to achieve $2\sigma$ consistency with our dynamical results. A combination of both corrections manages to produce results that are in agreement with our dynamical mass and radius within $1\sigma$.

\subsection{Surface gravity and average density}\label{sec:rhog}
The bottom two panels of Fig.~\ref{fig:mr_ast} show 1-dimensional plots of average density $\rho$ (in units of $10^{-3}\rho_{\odot}$) and surface gravity $\log g$ (dex) for the RG, both dynamical and asteroseismic measurements. Color and marker scheme is the same as for the upper two plots. 
We included the forward model by \citet{buldgen2019} by deriving from mass and radius, but chose not to propagate the uncertainties from them. \citet{buldgen2019} assumes a precision of $\sim 1.5\%$ on their inverted densities, which is approximately the same precision as the scaling relation uncertainties.

Considering the average density, we find that the uncorrected scaling relation underestimates density significantly, by almost $8\sigma$ when using the asteroseismic uncertainty, corresponding to $7\%$ lower than the dynamical value. Corrections are clearly needed when measuring $\rho$ from scaling relations for this system, which is in line with the findings of \citet{Brogaard2018} for three other systems. In \citetalias{Gaulme2016} the scatter of the observed $\rho_{\rm seis} - \rho_{\rm dyn}$ between the 10 different systems is too large to conclude if this trend is also present for their whole sample. Specifically, they find a mean $\rho_{\rm seis}$ lower than $\rho_{\rm dyn}$ by $2.5\%$ with a standard deviation of $4.7\%$. 

When using the $\Delta\nu$ correction by \citet{Rodrigues2017}, consistency between our asteroseismic and dynamical density is improved and mutual $1\sigma$ agreement is achieved. This is also in line with the findings of \citet{Brogaard2018} for three other EBs. As the sample of high-accuracy dynamical measurements of oscillating RGs increases, we will be able to affirm whether this theoretical correction is sufficient, or only an improvement relative to the uncorrected relations.


When comparing the $\log g$ measurements, we first conclude that deviation between dynamical and uncorrected asteroseismic $\log g$ is much smaller than for $\rho$. The uncorrected asteroseismic $\log g$ is $0.4\%$ larger than the dynamical result. The theoretical $\nu_{\rm max}$ correction from \citet{Viani2017} improves consistency slightly, allowing us to achieve $1\sigma$ agreement between asteroseismic and dynamical $\log g$. A larger sample would be needed to resolve whether a theoretical correction to the asteroseismic scaling relation of $\nu_{\rm max}$ is necessary.

The asteroseismic scaling relations for $\rho$ and $\log g$ (Eq.~\eqref{eq:01},~\eqref{eq:02}) only depend on one of the asteroseismic parameters each. We recommend that future works also focus on obtaining agreement between dynamical and asteroseismic $\rho$ and $\log g$ directly, instead of just mass and radius. Otherwise, parameter correlations could lead to false confirmations where simultaneous asteroseismic underestimation (or overestimation) of density and gravity produces consistent masses and radii. On the other hand, dynamical measurements of stellar radii are sensitive to third light, while masses are not for inclination close to $90^\circ$. In the presence of significant third light, the accuracy of dynamical $\rho$ and $\log g$ can thus be much lower than the accuracy on stellar mass. In such cases, if independent and accurate estimates of contamination are not obtained, the dynamically derived mass of the RG can still safely be compared with asteroseismology.
\subsection{Frequency modelling by \citet{buldgen2019} and \citet{Joergensen2020}}
Included in the Mass--Radius, density and surface gravity plots in Fig.~\ref{fig:mr_ast} are also results of a forward model of the individual oscillation frequencies by \citet{buldgen2019}, using only the metallicity and no other non-seismic constraints. They used the \emph{AIMS} modelling software \citep{rendle2019}, and a two-term surface correction from \citet{ballgizon2014}. This is intriguing in the context of this paper, as it matches the dynamical results very well. But the authors do caution that this was a preliminary model result, and that they used Eddington atmospheres that do not reproduce the effective temperature of RGB stars. As such, we will not assign strong weight to it. However, we can compare with the mean density inversion results that the \citet{buldgen2019} article focuses on. By assuming a precision of $\pm 1.5 \%$ as they do in \citet[Sect. 4.2]{buldgen2019}, all of their 4 inversion results, including their reference model for KIC\,8430105, agree with our dynamical density within $1\sigma$ of their uncertainties. 

\citet{Joergensen2020} performed similar modelling of the RG component in the system using effective temperature and metallicity from \citetalias{Gaulme2016}, along with either $\nu_{\rm max} = 76.70 \pm 0.57 \mu$Hz or $R = 7.65 \pm 0.05 R_{\odot}$ from \citetalias{Gaulme2016} as priors to obtain mass and radius. This could have biased their obtained results for all of their surface corrections. Only 2 out of 9 of their results with prior on $\nu_{\rm max}$ produce both mass and radius $1\sigma$ consistent with our dynamical measurements (S15a, S15b). Their internal uncertainty on average density for each surface correction (using prior on $\nu_{\rm max}$ and not on radius) is significantly smaller than our estimated uncertainty on the dynamical measurement in all but two cases, and their obtained average density varies by $0.6 - 6.7 \%$ from our dynamical result depending on their chosen surface correction (Andreas Christ Sølvsten Jørgensen, private comm.). For reference, our estimated $1\sigma$ uncertainty on the dynamical average density measurement is $0.9\%$. With a total of 9 models using different (or no) surface corrections and no prior on radius, 2 of their average density measurements vary by less than $1\%$ from our dynamical result (S15a, S15b), 4 vary by less than $2\%$ (\dittoclosing, BG14b, K08b), and 7 vary by less than $3\%$ (\dittoclosing, K08a, NoSC, BG14a). Overall, all of their models produce average densities, masses and radii that are an improvement over the uncorrected asteroseismic scaling relation using our measured $\Delta\nu_0$ (deviation $7.2\%$). Additionally, all the \citet{Joergensen2020} models with prior on $\nu_{\rm max}$ and deviation on average density of $<6\%$ produced densities that were lower than our dynamical measurement (7 of 9 models). Those last two models with large deviation also had significantly increased internal uncertainty on density ($2.5-3\%$).
\FloatBarrier
\section{Summary, conclusions and outlook}\label{sec:conclusion}
The SB2 eclipsing binary KIC\,8430105 has been analysed in detail with the purpose of comparing asteroseismic measurements of mass, radius, average density and surface gravity of its oscillating RG to results of our dynamical study using \emph{Kepler} photometry and spectroscopic RVs from the FIES instrument at the NOT. We obtained dynamical masses and radii that were more precise and significantly lower for both components than the previous study of \citetalias{Gaulme2016}, by $2.8$ and $3.5\sigma$ for the RG. These deviations were examined and the likely main cause is large epoch-to-epoch RV drifts in the spectroscopic follow-up by \citetalias{Gaulme2016} which reduced the quality of their obtained radial velocities relative to their reported RV uncertainties.

Since our asteroseismic measurements are similar to those of \citetalias{Gaulme2016}, the discrepancy between the scaling relation- and dynamical measurements are only increased by our analysis. The average density of the RG derived from the uncorrected asteroseismic scaling relation for $\Delta\nu$ (Eq.~\ref{eq:01}, $f_{\Delta\nu}=1$) was found to be $7\%$ lower than the dynamical measurement. This corresponded to a deviation of $\sim 8$ times the asteroseismic $1\sigma$ uncertainty. Similarly, mass and radius was significantly overestimated by $26$ and $11 \%$, respectively, when using the uncorrected scaling relations.

Despite finding dynamical mass and radius for the giant that were significantly lower than \citetalias{Gaulme2016}, we did manage to produce $2\sigma$ consistency between asteroseismic and dynamical mass and radius measurements if we applied a theoretical correction from \citet{Rodrigues2017} to the $\Delta\nu$ scaling relation. For the directly affected stellar parameter, the average density, this correction produced $1\sigma$ consistency between asteroseismic and dynamical measurements. These findings are consistent with the findings of Paper I for the mass of three other SB2 eclipsing binaries hosting an oscillating RG component.

Consistency between asteroseismic scaling relations and dynamical measurements is more clearly determined when using the independent relations, Eq.~\eqref{eq:01}~and~\eqref{eq:02}, instead of the mass and radius relations, Eq.~\eqref{eq:03}~and~\eqref{eq:04}. We therefore recommend that future works focus on obtaining agreement between asteroseismic and dynamical $\rho$ and $\log g$ before attempting to calibrate scaling relations to mass and radius. However, in cases where third light is significant and not well determined, where stellar radii, $\rho$ and $\log g$ could be affected, the dynamically measured mass can still be used as it is relatively insensitive to inclination changes when $i\sim 90^\circ$.

The $\nu_{\rm max}$ scaling relation correction by \citet{Viani2017} provided a small improvement to the asteroseismic $\log g$ measurement for the RG component in KIC\,8430105. With it, we managed to obtain $1\sigma$ consistency with the dynamically determined $\log g$, mass and radius of the giant. However, as uncorrected asteroseismic $\log g$ only deviated by $0.4\%$ from the dynamical measurement, it is not immediately obvious that a correction to $\nu_{\rm max}$ is warranted.

A model dependent age of the system was estimated from the metallicity, $[\alpha/\rm Fe]$ abundance, and dynamical mass and radius of the giant to be $3.7\pm0.4$Gyr.

The distance to the system was derived from Gaia EDR3 parallaxes and photometry to be $726.3 \pm 6.6$pc. This was then used with the spectroscopic temperature of the giant to produce a separate estimate for the radius of the giant of $7.71\pm 0.28R_{\odot}$, consistent with the dynamically determined radius of $7.475\pm0.031 R_{\odot}$.

Whether the modelled $\Delta\nu$ correction by \citet{Rodrigues2017} is sufficient at the current asteroseismic precision level or only an improvement should be examined through a large high-precision sample of SB2 eclipsing binaries hosting an oscillating giant component. If the full sample studied by both \citetalias{Gaulme2016} and \citet{Benbakoura2021} is reexamined with high precision radial velocity follow-up, potentially along with future SB2 detections from the \citet{gaulmeguzik2019} catalogue, we expect that the accuracy of the \citet{Rodrigues2017} and other $\Delta\nu$ corrections can be established. It should also be possible to resolve whether a correction to the $\nu_{\rm max}$ relation is required for RGs at any reasonable level of precision, and if so, if the theoretical correction by \citet{Viani2017} solves this need. 
%
%
\section*{Acknowledgements}
Based on observations made with the Nordic Optical Telescope, owned in collaboration by the University of Turku and Aarhus University, and operated jointly by Aarhus University, the University of Turku and the University of Oslo, representing Denmark, Finland and Norway, the University of Iceland and Stockholm University at the Observatorio del Roque de los Muchachos, La Palma, Spain, of the Instituto de Astrofisica de Canarias.

We thank the referee Patrick Gaulme for a detailed and constructive report that helped improve the paper.

JST thanks Rasmus Handberg for in-depth and enlightening discussion throughout the preparation of this paper.

Funding for the Stellar Astrophysics Centre is provided by The Danish National Research Foundation (Grant agreement no.: DNRF106). 


This paper includes data collected by the Kepler mission. Funding for the Kepler mission is provided by the NASA Science Mission directorate.

This paper includes data collected by the TESS mission. Funding for the TESS mission is provided by the NASA's Science Mission Directorate.

AM acknowledges support from the ERC Consolidator Grant funding scheme (project ASTEROCHRONOMETRY, \url{https://www.asterochronometry.eu}, G.A. n. 772293).

This research makes use of public auxiliary data provided by ESA/Gaia/DPAC/CU5 and prepared by Carine Babusiaux.

Some of the data presented in this paper were obtained from the Mikulski Archive for Space Telescopes (MAST). STScI is operated by the Association of Universities for Research in Astronomy, Inc., under NASA contract NAS5-26555. Support for MAST for non-HST data is provided by the NASA Office of Space Science via grant NNX09AF08G and by other grants and contracts.

This work has made use of data from the European Space Agency (ESA) mission
{\it Gaia} (\url{https://www.cosmos.esa.int/gaia}), processed by the {\it Gaia}
Data Processing and Analysis Consortium (DPAC,
\url{https://www.cosmos.esa.int/web/gaia/dpac/consortium}). Funding for the DPAC
has been provided by national institutions, in particular the institutions
participating in the {\it Gaia} Multilateral Agreement.

This research made use of Lightkurve, a Python package for Kepler and TESS data analysis \citep{lightkurve2018}.

This research made use of Astropy (\url{http://www.astropy.org}) a community-developed core Python package for Astronomy \citep{astropy:2013, astropy:2018}. 
\section*{Data Availability}
All data underlying this article are available directly in the text or through the references cited, with a few exceptions: Measured oscillation frequencies, separated component spectra, FIES@NOT spectroscopic follow-up observations, and intermediate data products (uncertainty estimation sample fits, normalized and clipped light curves), which will be shared upon reasonable request.




\bibliographystyle{mnras}
\bibliography{thomsen}




\FloatBarrier
\newpage
\appendix
\section{Table with RV measurements}
\begin{table*}
\centering
\caption{Spectroscopic observation summary, including radial velocity measurements (RV), S/N, and weight given when producing separated component spectra.}
\label{tab:rv8430105} 
\begin{tabular}{lcccccc}
\hline
Phase & BJD-2450000 & $\rm RV_G$ [km/s] & $\rm RV_{MS}$ [km/s] & S/N@(5605-5612\AA) & Weight & Within eclipse \\
\hline
0.134041    & 8426.38576    & 0.635(51)     & 29.63(54)      & 30  &  1 &  \\
0.166301    & 8428.42866    & -2.1005 (86)  & 32.97(42)      & 26  &  1 & \\
0.181986    & 8429.42195    & -3.255(34)    & 33.94(65)      & 25  &  0 &  \\
0.307002    & 8437.33884    & -8.627(39)    & 43.84(59)      & 26  &  1 &  \\
0.434258    & 9078.66868    & -7.482(49)    & 42.22(38)      & 31  &  1 &  \\
0.449166    & 8446.34171    & -6.894(55)    & 40.6(16)      & 22  &  0 &  \\
0.489235    & 8385.55207    & -4.622(35)    & 35.38(55)      & 25  &  1 &  \\
0.503868    & 8386.47874    & -3.518(35)    & 35.40(41)      & 27  &  1 &  \\
0.575423    & 8454.33717    & 4.113(47)     & 23.57(40)      & 29  &  0 &  \\
0.584057    & 8391.55685    & 5.324(37)     & 22.31(60)      & 32  &  0 &   \\
0.598637    & 8392.48014    & 7.448(32)     & 17.93(46)      & 26  &  0 &   \\
0.614287    & 8393.47123    & 10.073(52)    & $\sim$      & 29  &  0 &   \\
0.644563    & 8395.38852    & 15.667(63)    & $\sim$      & 23  &  0 & True  \\
0.645394    & 8395.44116    & 15.870(40)    & $\sim$      & 27  &  0 & True  \\
0.654123    & 8459.32106    & 17.547(43)    & $\sim$      & 13  &  0 & True  \\
0.660421    & 8396.39278    & 18.953(43)    & $\sim$      & 21  &  0 & True  \\
0.793954    & 9101.44719    & 45.187(45)    & -39.06(45)      & 21  &  1 &   \\
0.811082    & 9102.53187    & 46.218(43)    & -41.71(31)      & 20  &  1 &   \\
0.835652    & 8407.48963    & 45.866(47)    & -39.7(11)      & 31  &  1 &   \\
0.873724    & 9106.49879    & 41.969(49)    & -34.89(49)      & 25  &  1 &   \\
0.972876    & 9429.42027    & 23.481(38)    & -5.61(50)      & 27  &  1 &   \\
\hline
\end{tabular}
\end{table*}
\section{Deformation and reflection}\label{sec:forward}
\begin{figure}
    \centering
    \includegraphics[width=\columnwidth]{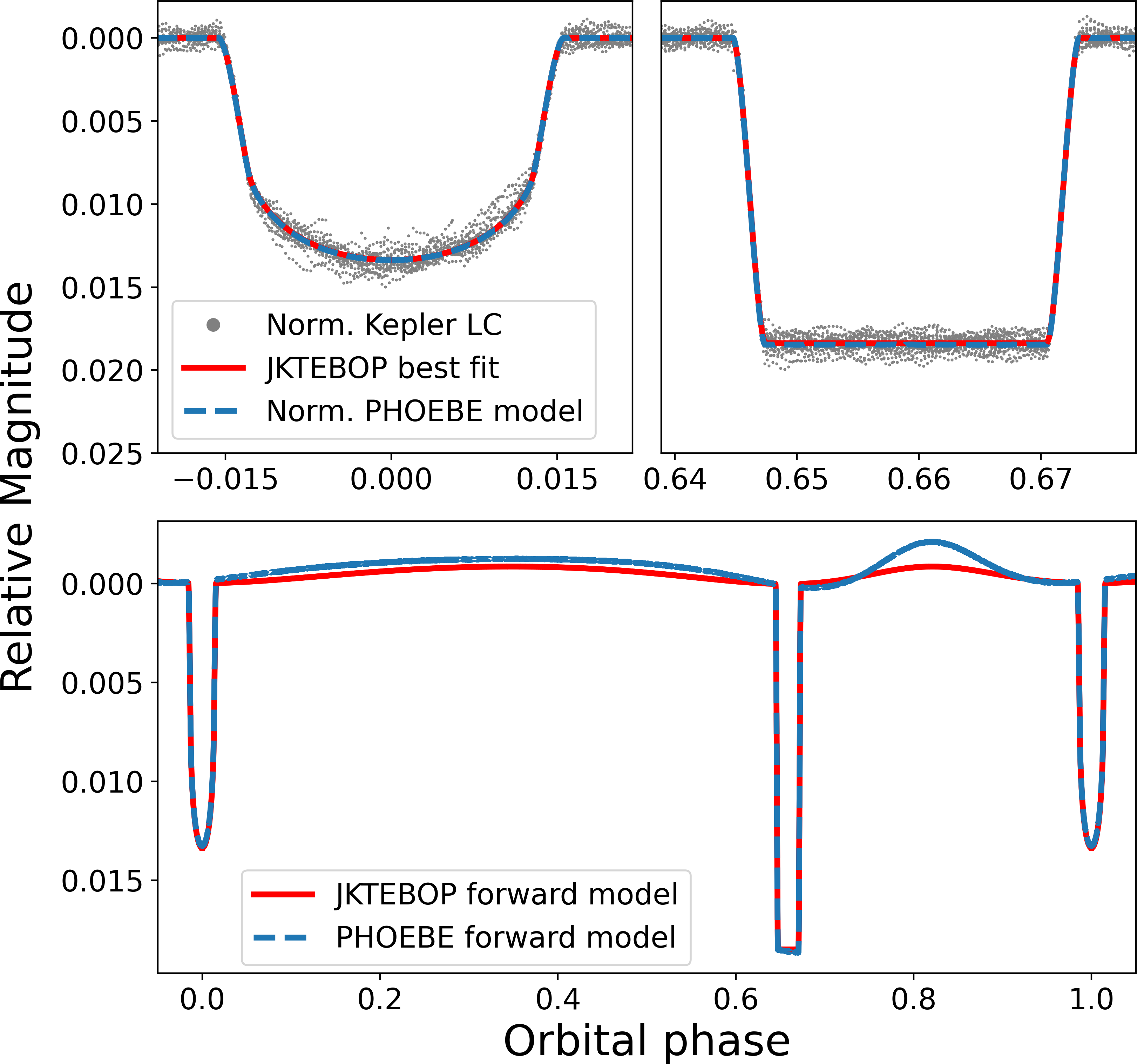}
    \caption{Top two panels: Normalized PDCSAP \emph{Kepler} light curve (grey markers), best JKTEBOP fit without deformation or reflection (red), and a normalized PHOEBE forward model with deformation and reflection (blue). Bottom panel: A rescaled PHOEBE forward model (blue), and a rescaled JKTEBOP forward model with deformation and reflection (red). Both forward models use parameters from the best-fit \emph{Kepler} solution in Table~\ref{table:EBdata}.}
    \label{fig:eb_forward}
\end{figure}

To examine deformation and reflection modelling in the system, we compared with the more advanced models of the PHOEBE 2 code \citep{conroy2020} for the \emph{Kepler} passband. The models were generated using the parameters obtained from eclipsing binary analysis of the normalized PDCSAP \emph{Kepler} light curve with JKTEBOP, shown in Table~\ref{table:EBdata}. The lower panel in Fig.~\ref{fig:eb_forward} compares a JKTEBOP model using the bi-axial ellipsoid treatment of deformation and reflection outside eclipses (red) and the PHOEBE model that uses a Roche treatment for deformation and Wilson treatment of reflection (blue) \citep{Wilson1990}. The magnitude zero-point of the JKTEBOP and PHOEBE models have been independently re-scaled. Unsurprisingly, the JKTEBOP model does not reproduce the PHOEBE model outside of eclipses. The system geometry is such that the PHOEBE model indicates a non-flat light curve inside the total eclipse of the MS star. 

The upper panel in Fig.~\ref{fig:eb_forward} compares the normalized \emph{Kepler} data (grey dots) with the best-fit JKTEBOP solution without deformation and reflection (red) and the PHOEBE forward model normalized with polynomial fitting (blue). By normalizing the PHOEBE forward model and truncating it to only cover the eclipses as with the observed \emph{Kepler} light curve for eclipsing binary analysis, we can compare a JKTEBOP solution obtained using the PHOEBE model in place of the light curve, with photometric uncertainty estimate from the \emph{Kepler} light curve. This results in a radius of the giant component lower by $0.1$ times our estimated $1\sigma$ uncertainty, radius of the MS component lower by $0.04\sigma$, and an inclination of $89.59^\circ$, $0.1\sigma$ larger than previous. Some of this variation could be caused by differences in how JKTEBOP defines the luminosity ratio relative to PHOEBE fluxes, increase in the apparent photometric precision of the PHOEBE model compared with the PDCSAP light curve, lack of stellar oscillations in the PHOEBE model, or a lower observed radii during eclipses due to deformation. Overall this verifies that polynomial fitting around the eclipses is unlikely to have introduced a significant bias in stellar parameters when removing effects of deformation and reflection. We have not investigated whether the method is equally successful for removing long-term variability such as magnetic activity and instrumental effects. However, since the timescale for these effects is similar or longer, we expect this normalization method to perform just as well.
\section{Third light}\label{sec:tlight}
Our only estimate of contamination for the KASOC light curve is from the standard \emph{Kepler} photometric pipeline, which uses a different aperture. We compared with results obtained using the PDCSAP processed light curve. Most parameters were unaffected by this. Importantly, we observed no change in the found stellar masses. However, the inclination, stellar radii, and through this the density and $\log g$, all changed by approximately $1\sigma$ (increased inclination, decreased radii).

We attempted to fit for third light in JKTEBOP while using the KASOC light curve. This resulted in a fitted third light fraction of $7\%$ with an internal uncertainty of $\pm 3\%$ from the minimizer. This is on par with the TESS pipeline estimate, and much higher than the \emph{Kepler} pipeline mean estimate of $0.3\%$. The resulting radius of the giant was $7.454 R_{\odot}$, now consistent with the PDCSAP derived radius. However, the inclination was $89.90^{\circ}$, and the radius of the MS star was $0.7710 R_{\odot}$, almost $5\sigma$ higher than when using the PDCSAP light curve without fitting for third light. JKTEBOP throws a warning for inclinations above $89.90^{\circ}$, since the program forces the minimizer to find solutions below $90^{\circ}$ and can therefore tend to skew results. This could have caused the minimizer to favor a biased inclination coupled with changes in third light and radii instead. This is discussed in more detail in Appendix.~\ref{sec:uncertainty}.

We inspected the \emph{Kepler} and TESS target pixel files using Lightkurve \citep{lightkurve2018} along with stellar catalogues to determine whether there was an expected source of contamination that could explain this. A $13$ magnitude star KIC\,8430119 is on the edge of the \emph{Kepler} target pixel files for KIC\,8430105. It is the only other star with magnitude $<15.5$ within the TESS pipeline aperture. It is possible that the KASOC light curve includes a higher number of pixels containing light from this star than the \emph{Kepler} pipeline aperture. This could lead to a contamination of up to $\sim 10\%$ at absolute maximum from just this star. Since the KASOC aperture and background masks are not available, it was not possible to confirm whether this has occurred.

It was observed that for some of the \emph{Kepler} quarters, the \emph{Kepler} pipeline aperture mask would extend slightly nearer KIC\,8430119. It correlated well with simultaneous variations in the individual quarter pipeline contamination estimates of $0.09\%$ - $0.5\%$. However, whether or not those estimates were over/underestimated as a whole was not examined in detail. Since KIC\,8430119 was the only detected source within the target pixel file, we found it reasonable to assume that these estimates were accurate at least up to $100\%$ the size of the reported values.

We also examined the effect of fitting for third light in the PDCSAP \emph{Kepler} light curve. This resulted in a fitted third light fraction of $2.6 \%$ with internal uncertainty $2.8\%$. We noticed a similar trend as with the KASOC light curve, with the resulting radius of the RG being $7.451 R_{\odot}$, inclination $89.95^{\circ}$, and increased radius of the MS star of $0.7560 R_{\odot}$. To investigate the behavior of JKTEBOP in this case, a static contamination of $10\%$ was introduced by decreasing the depth of the eclipses with $\rm  flux_* = 0.9\cdot flux + 0.1$. As expected, third light was then found to be $12.6 \pm 2.7\%$, with same results as before.

From all of this, we conclude that in order to produce highly accurate eclipsing binary measurements for benchmarking of other methods, it is essential that the photometric pipeline used is able to independently estimate contamination in a reliable way, as the eclipsing binary analysis might not be able to account for it properly at the needed precision level. This is supported by \citet{Southworth2010}. Since the KASOC pipeline does not currently provide contamination estimates and the original processing masks were not stored, which would make it possible to after-produce them, we only report stellar parameters produced using the PDCSAP version of the \emph{Kepler} light curve in Table~\ref{table:EBdata}. 

As we were not able to produce reliable fits while fitting freely for third light in the PDCSAP light curve, we decided to use the PDCSAP light curve with a fixed third light fraction of 0.0 in order to produce the best fit. Third light was then a free parameter during uncertainty estimates. Attempting this with no prior on third light gave a confidence interval of $[-5, 2]\%$ for this parameter, a $\gtrsim 100\%$ increase in uncertainties of correlated parameters (radii, $\rho$, $\log g$ and inclination) and clearly skewed corner plots of correlated fitting parameters. Most solutions were found with inclination $\sim 90^{\circ}$. Given that this comes from an inability to properly fit for third light, and since we have reasonable arguments to assume that the PDCSAP light curve does not have such a high degree of third light unaccounted for, we did not find that this increase properly reflects the actual uncertainty present. It was therefore decided to include a Gaussian prior of $0.000\pm0.005$ for the third light fraction when producing parameter uncertainty estimates through the JKTEBOP option to add a third light measurement (THDL). The uncertainty here was chosen to be the maximum reported contamination of all the quarters in the \emph{Kepler} SAP light curve. This is a very conservative estimate since it indicates that the \emph{Kepler} pipeline could be off by at minimum $100\%$ the reported value. When producing uncertainty estimates with a fixed third light of $0.0$ instead, we found no observable change in uncertainties or corner plots for the \emph{Kepler} PDCSAP light curve. The third light estimate would only be off by more than $100\%$ if it did not properly account for KIC\,8430119, or if a very dim third companion was present. Only the tail of KIC\,8430119 was within the aperture pixels, and we therefore consider it unlikely that significant contamination is present from this star. However, giving an uncertain prior allows us to be absolutely sure that the high precision measurements we report in Table~\ref{table:EBdata} are, in fact, not biased due to third light. The only other possible contaminating factor than KIC\,8430119 would be a third component to KIC\,8430105. We did not observe any spectroscopic signal for such a companion. It would therefore have to be significantly dimmer than the MS component, meaning $\ll 2\%$ of the light. Its presence would be unlikely to have an impact on the parameters found.

The TESS light curve had an estimated contamination of $7.6\%$, much larger than for the \emph{Kepler} aperture. A conservative prior like with the \emph{Kepler} light curve is not feasible as we become limited by the optimizer near $i\sim 90^\circ$ (see Appendix~\ref{sec:uncertainty}). We have not evaluated the accuracy of the estimate, and would therefore prefer to keep an uncertain prior still. We chose a compromise with uncertainty slightly above $50\%$. Contamination from KIC\,8430119 is not above $10\%$, and other stars within the aperture are $\sim2.5$mag dimmer than this. A prior of $7.6\% \pm 4.0\%$ would therefore ensure that we could potentially have contamination up to $\sim 12\%$. The large contamination present, and our preference to an uncertain prior on third light, is one of the reasons that we do not assign any significance to the stellar parameters obtained from eclipsing binary analysis with the TESS light curve.

During our uncertainty estimation in Appendix.~\ref{sec:uncertainty}, we concluded that fixing the third light to $7.6\%$ while finding the best fit produced a biased measurement of inclination. Therefore, for the TESS light curve we chose to report the median solution from the residual block bootstrap in Table~\ref{table:EBdata} instead of the best fit to the original light curve. Both the median solution and the best fit are in overall agreement with the results produced with the \emph{Kepler} PDCSAP light curve, up to the lower precision afforded by the TESS light curve.

It would benefit studies of this nature greatly if public tools were developed for astronomers to convert ground-based follow-up photometry to expected observations in \emph{Kepler} pixels in a simple fashion. This would enable us to perform our own per-cadence contamination estimates and contamination uncertainty estimates, thereby securing a higher degree of confidence in obtained results than can be done with what is currently available. A new mission similar to \emph{Kepler} with observation length, photometric precision and pixel size allowing this degree of quality is not in the works and unlikely to be for many years (with HAYDN \citep{Miglio2021b} a potential exception). Therefore, we consider this to be one of the key improvements that could be made for the use of eclipsing binaries as a benchmarking tool for $\sim 1\%$ precision model-dependent measurements such as asteroseismology.
\section{Uncertainty estimation}\label{sec:uncertainty}
The residual block bootstrap used to calculate the uncertainties of the dynamical parameters in Table~\ref{table:EBdata} is a non-parametric uncertainty estimation method. It divides the best-fit light curve model produced using JKTEBOP into a set of static non-overlapping blocks, preserving the original timestamps of the data. It then samples new synthetic light curves by assigning a random block of residuals of the same size to each non-overlapping model block. The residual blocks are chosen independently of the model blocks and are allowed to overlap. This makes the method somewhat similar to a moving block bootstrap without changing the orbital phase-coverage. We did not have a decent argument for choosing a specific block-length. Therefore, we decided to perform the residual block bootstrap multiple times with different block-lengths. For the \emph{Kepler} light curve, chosen block-lengths correspond to [1, 1/2, 1/3, 1/4, 1/5, 1/6, 1/7, 1/8] the length of each eclipse, and for the TESS light curve this was instead [1/2, 1/4, 1/6, 1/8, 1/10, 1/12, 1/14, 1/16] due to the higher cadence allowing smaller block sizes.

The radial velocity observations are assumed to be uncorrelated in time and phase, and synthetic RVs are randomly resampled from the residuals and best fit. We preserved the original timestamps as well, just not any ordering of the residuals.

While computed 1-dimensional standard deviations did vary slightly with the different choices of block-lengths for the light curve, we did not observe any specific correlation with block-length or between the parameters. By comparing corner-plots produced with each block-length independently, we found distributions that were very consistent with each other. Therefore we collapsed all attained results for all block-lengths into one when calculating final uncertainties.

We compared the residual block bootstrap with two non-parametric methods from JKTEBOP, the residual permutation method \citep[TASK 9, ][]{Southworth2008} and a Monte Carlo simulation \citep[TASK 8, ][]{Southworth2004}. $1\sigma$ uncertainties found with each when using our PDCSAP \emph{Kepler} light curve and our NOT RVs are shown in Table~\ref{table:uncertainty} in the appendix. Overall, with the PDCSAP \emph{Kepler} light curve and NOT RVs, TASK 9 produced uncertainty estimates which were slightly larger than our residual block bootstrap on average. In comparison, TASK 8 produced estimates which were consistently and significantly lower than the residual block bootstrap for almost all parameters. However, when using the \citetalias{Gaulme2016} RVs (with a static error addition of $0.88\rm km s^{-1}$ to the RVs of the giant), TASK 9 produced estimates which were even lower than both the Monte Carlo simulation and the residual block bootstrap and clearly underestimated. This is not unexpected, since the \citetalias{Gaulme2016} RVs include a large spread, likely from epoch-to-epoch calibration drifts, which is not clearly time-correlated. Currently, it is suggested to use TASK 8 in such cases instead. However, this method neglects any correlated noise present in the light curve. In that case the Monte Carlo simulation produced estimates that were still underestimated compared with the residual block bootstrap.

When third light was a free or badly constrained parameter, the minimizer had a tendency to favor adjusting it and compensating with the inclination and sum of the relative radii, until it reached $i\sim 90^{\circ}$ and was locked into a local minimum. We investigated if this stems from one of the shortcuts in the way that JKTEBOP numerically handles inclinations above $90^{\circ}$, and found that even if we disabled the symmetric reflection of EBOP ($91^{\circ}=89^{\circ}$) and disabled the subsequent devaluing of solutions with $i > 89.9^{\circ}$ by JKTEBOP, this effect was still present. 

We consider this to be the most significant limitation of a non-parametric uncertainty estimation method; it is limited by the optimizer used in the fitting algorithm. The method is not sampling the parameter space semi-randomly to estimate the likelihood of different solutions, but instead resampling the data with the assumption that the optimizer is unbiased. If the fitting algorithm consistently favours a specific configuration of parameters due to a kink in the parameter space, that information is lost to us afterwards. It is therefore not a guarantee that we will be able to freely marginalize over parameters that we have no independent information on in the light curve. However, our experience here is that this issue would be avoided for either slightly lower inclinations, third light $\lesssim1\%$, or if an accurate and precise independent estimate of third light is available.

When constraining the third light to a reasonable physical uncertainty in the \emph{Kepler} light curve, we mostly overcome this since only a limited number of fits favour inclinations above $89.9^{\circ}$. We still find a clear bi-modal solution for the inclination, but fixing third light to 0.0 does not remove this. It is primarily the sum of relative radii that would be significantly changed if we adopted the solution with $i \sim 90^{\circ}$. As a result, density, surface gravity and radius would change by just under $1\sigma$ for both components. This uncertainty is reflected in the reported confidence intervals in Table~\ref{table:EBdata}.

For the TESS light curve, the median solution and best fit do not agree as well as with the \emph{Kepler} light curve, especially for inclination and third light. The best fit found $i>89.9^{\circ}$, which was just argued as unlikely to be the true value. The inclination distribution showed the multi-modal trends also observed with the \emph{Kepler} light curve to a much higher degree. This is due to a badly constrained third light, which we only estimated a prior of $7.6\%\pm4.0\%$ for in Sect.~\ref{sec:tlight}. While the median solution inclination is likely biased due to this as well, we expect that it is closer to the true value. Therefore, the median solution (0.5 quantile) is reported in Table~\ref{table:EBdata} for the TESS light curve, instead of the best fit.

With the shortcomings of our non-parametric method in mind, it is also worth summarizing some of its merits: The main takeaway is its simplicity, both in functionality and in implementation. The ability to produce relatively reliable uncertainty estimates in the presence of un-modelled signal shows this. By this we mean that it is possible to approximate undesired astrophysical effects as a form of correlated noise when resampling from the data, which severely reduces model complexity in the long run. In a parametric sampling method it would have been necessary to marginalize over all such effects in the model. Simplicity of implementation comes from the fact that we could build this on top of an already implemented solver like JKTEBOP. The only two ingredients needed were a way to resample light curves and RVs, and an Input/Output interface for the JKTEBOP program.
\begin{table*}
\centering
\caption{\label{table:uncertainty} $1\sigma$ uncertainty estimates for dynamical parameters with PDCSAP \emph{Kepler} light curve and NOT RVs.}
\begin{threeparttable}
\begin{tabular}{lccc}
\hline
\hline
Quantity & RBB\tnote{a} & TASK 9\tnote{b} & TASK 8\tnote{c} \\
\hline
$P$ ($10^{-5}$days)                                          & $6.6$  & $7.2$ & $2.9$ \\
$\rm t_{\rm RG}$  ($10^{-3}$days)                            & $1.6$   & $1.6$ & $0.64$ \\
$i$ ($\circ$)                                       & $0.22$     &  $0.27$   & $0.19$ \\
$e$ ($10^{-4}$)                                                & $1.0$    & $1.4$ & $0.53$    \\
$\omega$ ($\circ$)                                  & $0.12$ & $0.17$     & $0.067$      \\
$e\cos\omega$ ($10^{-5}$)                                       & $4.1$ & $3.9 $       & $1.7$ \\
$e\sin\omega$ ($10^{-4}$)                                      & $5.4$ & $7.8$      & $3.0$ \\
$r_{\rm MS}+r_{\rm RG}$ ($10^{-4}$)                            & $2.8 $ & $3.2$    & $2.4 $  \\
$k = r_{\rm MS}/r_{\rm RG}$ ($10^{-4}$)                         & $2.8$& $3.0$    & $2.5$   \\
$J$ ($10^{-3}$)                                                 & $8.1$ & $7.5$      & $3.1$     \\
$\frac{L_{\rm MS}}{L_{\rm RG}}$ ($10^{-5}$)\tnote{d}            & $5.4$ & $6.0$    & $8.5$  \\
$K_{\rm RG}$ (km/s)                                     & $0.013$ & $0.014$     & $0.013$    \\
$K_{\rm MS}$ (km/s)                                 & $0.22$ & $0.27$      & $0.15$      \\
$a (R_{\odot})$                                     & $0.26$ & $0.32$     & $0.19$      \\
$\gamma_{\rm RG}$ (m/s)                                  & $9.6$ & $14$    & $7.4$     \\
$\gamma_{\rm MS}$ (km/s)                                 & $0.16$ & $0.13$     & $0.12$     \\
Mass$_{\rm RG}(M_{\odot})$                          & $0.014$ & $0.017$      & $0.010$      \\
Mass$_{\rm MS}$ ($10^{-3}M_{\odot})$                          & $5.1$ & $6.2$      & $3.6$      \\
Radius$_{\rm RG}(R_{\odot})$                        & $0.031$ & $0.036$     & $0.024$     \\
Radius$_{\rm MS}$ ($10^{-3}R_{\odot})$                        & $4.1$ & $4.3$      & $3.1$      \\
log$g_{\rm RG}$ ($10^{-3}$)                              & $3.3$ & $3.9$     & $2.5$       \\
log$g_{\rm MS}$ ($10^{-3}$)                               & $3.9$ & $3.8$     & $3.0$       \\
$\rho_{\rm RG} (10^{-3}\rho_{\bigodot})$            & $0.026$ & $0.030$     & $0.022$                   \\
$\rho_{\rm MS} (\rho_{\bigodot})$                   & $0.026$ & $0.026$     & $0.021$                  \\
ld$_{a, \rm RG}$ ($10^{-3}$)                                   & $13$ & $13$      & $5.4$\\
$L_3$ ($\%$)                                    & $0.12$ & $0.15$ & $0.47$ \\
\hline
\end{tabular}

\begin{tablenotes}
		\scriptsize
\item[a] Our residual block bootstrap.
\item[b] JKTEBOP residual permutation algorithm.
\item[c] JKTEBOP Monte Carlo simulation.
\item[d] In the $K_p$ bandpass.

\end{tablenotes}
\end{threeparttable}
\end{table*}


\bsp	
\label{lastpage}
\end{document}